\documentclass[preprint]{aastex}
\author{Anastasia~Tsvetkova\altaffilmark{1,a}, 
	Dmitry~Frederiks\altaffilmark{1,b}, 
	Dmitry~Svinkin\altaffilmark{1}, 
	Rafail~Aptekar\altaffilmark{1,\dag},
	Thomas~L.~Cline\altaffilmark{2,*}, 
	Sergei~Golenetskii\altaffilmark{1},
	Kevin~Hurley\altaffilmark{3},
	Alexandra~Lysenko\altaffilmark{1}, 
	Anna~Ridnaia\altaffilmark{1}, 
	Mikhail~Ulanov\altaffilmark{1}
}
\altaffiltext{1}{Ioffe Institute, Politekhnicheskaya 26, St.~Petersburg 194021, Russia; \href{mailto:tsvetkova@mail.ioffe.ru}{tsvetkova@mail.ioffe.ru}}
\altaffiltext{2}{NASA Goddard Space Flight Center, Greenbelt, MD 20771, USA}
\altaffiltext{3}{University of California, Berkeley, Space Sciences Laboratory, 7 Gauss Way, Berkeley, CA 94720-7450, USA}
\altaffiltext{a}{\texttt{tsvetkova@mail.ioffe.ru}}
\altaffiltext{b}{\texttt{fred@mail.ioffe.ru}}
\altaffiltext{\dag}{Deceased}
\altaffiltext{*}{Retired}
\usepackage{cmap} 
\usepackage[english]{babel}
\usepackage{indentfirst}
\usepackage{amssymb}
\usepackage[unicode=true,draft]{hyperref}
\usepackage{pdflscape,natbib,color}
\usepackage{graphicx}
\usepackage{gensymb}
\usepackage{amsmath}
\usepackage{hyperref}
\slugcomment{\small{Accepted by ApJ}}

\title{THE KONUS-\textit{WIND} CATALOG OF GAMMA-RAY BURSTS WITH KNOWN REDSHIFTS. II. WAITING MODE BURSTS SIMULTANEOUSLY DETECTED BY \textit{SWIFT}/BAT.}
\begin{document}
\newcounter{cit}
\newcommand{\cititem}[1]{\refstepcounter{cit}(\arabic{cit})~\label{Gen:#1}\citealt{#1}}
\newcommand{\KW}{Konus-\textit{Wind} }
\newcommand{\BAT}{\textit{Swift}/BAT }

\begin{abstract}
In the Second part of The Konus-Wind Catalog of Gamma-Ray Bursts with Known Redshifts
(first part: \citealt{Tsvetkova2017}; T17), we present the results of a systematic study of gamma-ray bursts (GRBs) 
with reliable redshift estimates detected simultaneously by the \KW (KW) experiment (in the waiting mode)
and by the \BAT (BAT) telescope during the period from 2005~January to the end of 2018. 
By taking advantage of the high sensitivity of BAT and the wide spectral band of KW
we were able to constrain the peak spectral energies, the broadband energy fluences, and the peak fluxes 
for the joint KW-BAT sample of 167 weak, relatively soft GRBs (including four short bursts).
Based on the GRB redshifts, which span the range  $0.04 \leq z \leq 9.4$, we estimate the rest-frame, 
isotropic-equivalent energy, and peak luminosity. 
For 14 GRBs with reasonably constrained jet breaks, we provide the collimation-corrected values of the energetics. 
This work extends the sample of KW GRBs with known redshifts to 338 GRBs, the largest set of cosmological GRBs studied to date over a broad energy band. 
With the full KW sample, accounting for the instrumental bias, we explore GRB rest-frame properties, including hardness-intensity correlations, 
GRB luminosity evolution, luminosity and isotropic-energy functions, and the evolution of the GRB formation rate, 
which we find to be in general agreement with those reported in T17 and other previous studies. 

\end{abstract}
\section{Introduction}
\label{Introduction}
Gamma-ray bursts (GRBs) were discovered more than a half a century ago, 
and  their cosmological origin was established about two decades ago 
via identifications of the burst optical counterparts and their redshift measurements.
As of 2019, cosmological redshift ($z$) estimates are known for $\sim 500$ GRBs, 
ranging from spectroscopic $z = 0.0087$ (GRB~980425; \citealt{Foley2006}) to photometric $z = 9.4$ (GRB~090429B; \citealt{Cucchiara2011}) 
or NIR spectroscopic $z = 8.1$ (GRB~090423; \citealt{Salvaterra2009}). 

With the GRB redshift known, it is possible to estimate the isotropic-equivalent 
burst energetics and transform the GRB spectral parameters and durations from the observer to the cosmological rest frame.
At cosmological distances, huge isotropic-equivalent energy releases up to $E_\textrm{iso} \lesssim 10^{55}$~erg 
(e.g. GRB~080916C, \citealt{Abdo2009,Greiner2009a}) and isotropic peak luminosities 
$L_\textrm{iso} \lesssim 5 \times 10^{54}$~erg~s$^{-1}$ (e.g. GRB 110918A, \citealt{Frederiks2013}) 
can be explained under the assumption that GRBs are non-spherical explosions: 
when the tightly collimated relativistic fireball is decelerated by the circumburst medium (CBM) 
down to the Lorentz factor $\Gamma \sim 1/\theta_\textrm{jet}$ ($\theta_\textrm{jet}$ is the jet opening angle), 
an achromatic break (jet break) should appear, in the form of a sudden
steepening in the GRB afterglow light curve, at a characteristic time $t_\textrm{jet}$.
Correction for the jet collimation decreases the energy releases and peak luminosities of GRBs by orders of magnitude.
GRBs with known $z$ may be used as cosmological probes, e.g., to study the expansion rate of the Universe and 
to investigate the observational properties of dark energy if the empirical correlations between 
spectral properties and energy are appropriately calibrated.
GRB redshifts are also the keys to population studies: e.g., GRBs with high redshifts 
and their afterglows can serve as a unique probe of the first stars in the Universe. 

The \KW (hereafter KW, \citealt{Aptekar1995}) experiment has operated since 1994 November and plays 
an important role in GRB studies thanks to its unique set of characteristics: 
the spacecraft orbit in interplanetary space that provides an exceptionally stable background; 
the continuous coverage of the full sky by two omnidirectional detectors; 
and the broad energy range ($\sim$20~keV--15~MeV, triggered mode; $\sim$20~keV--1.5~MeV, waiting mode).
Since the KW energy band is rather wide, the GRB peak energy $E_\textrm{p}$ (the maximum of the $E F_{E}$ spectrum) can be derived directly from the KW spectral data
and the GRB energetics can be estimated using fewer extrapolations. 
This work extends the first part of the KW catalog of GRBs with known redshifts (\citealt{Tsvetkova2017}; T17) 
which provided temporal and spectral parameters, as well as broadband observer- and rest-frame energetics of 150 bursts 
detected in the KW triggered mode,  
with a sample-wide peak flux limit of $\sim 1 \times 10^{-6}$~erg~cm$^{-2}$~s$^{-1}$. 
Updated up to the end of 2018, this sample comprises 171 GRBs with $0.1 \leq z \leq 5$.

The Neil Gehrels \textit{Swift} Observatory, operating since 2004 November, is the primary source
of precise GRB localizations allowing for subsequent optical identifications of the sources and their redshift measurements. 
Thanks to the high sensitivity of the \textit{Swift}/Burst Alert Telescope (BAT; \citealt{Barthelmy2005}) telescope ($\sim10^{-8}$~erg~cm$^{-2}$~s$^{-1}$),
the BAT sample of GRBs with known $z$, which comprises $\sim 380$ events to date, is the largest one. 
However, the relatively soft spectral range of the instrument (15--150~keV) complicates GRB studies, 
since only for a limited fraction of soft-spectrum bursts can $E_\textrm{p}$ and broadband fluences be constrained from the BAT data.
Moreover, as \citet{Sakamoto2009} showed, even if the peak energy is located inside the BAT energy range, BAT still cannot derive $E_\textrm{p}$ if the burst is not bright enough.

Here, we present a sample of 167 GRBs with reliably measured redshifts detected by KW in the waiting mode and, 
simultaneously, observed by \textit{Swift}/BAT, from 2005 January to 2018 December.  
We start with a brief description of the KW and \BAT instruments in Section~\ref{Instrumentation}.
The burst sample is described in Section~\ref{Sample}. 
In Section~\ref{Analysis} we present the joint temporal and spectral analysis of KW and BAT data and derive the burst energetics.
In Section~\ref{Discussion} we discuss the analysis results, and the rest-frame properties of the whole sample of 338 KW GRBs with known $z$,
the largest set of cosmological GRBs studied to date over a broad energy band.
Our conclusions are summarized in Section~\ref{Conclusions}.

All the errors quoted in this catalog are at the 68\% confidence level (CL) and are of statistical nature only.
Throughout the paper, we assume the standard $\Lambda$CDM model: $H_0 = 67.3$~km~s$^{-1}$~Mpc$^{-1}$, $\Omega_{\Lambda} = 0.685$, and $\Omega_M = 0.315$ \citep{Planck2014}. 
We also adopt the conventional notation $Q_k = Q/10^k$.

\section{Instrumentation}
\label{Instrumentation}
\subsection{\KW}
KW is a gamma-ray spectrometer designed to study temporal and spectral characteristics of GRBs, solar flares (SFs), soft gamma-repeaters (SGRs), 
and other transient phenomena over a wide energy range from 13~keV to 10~MeV, nominally (i.e., at launch).  
It consists of two identical omnidirectional NaI(Tl) detectors, mounted on opposite faces of the rotationally stabilized \textit{Wind} spacecraft. 
One detector (S1) points toward the south ecliptic pole, thereby observing the south ecliptic hemisphere; the other (S2) observes the north ecliptic hemisphere. Each detector has an effective area of $\sim$80--160~$\rm cm^2$, depending on the photon energy and incident angle.
In interplanetary space far outside the Earth's magnetosphere, KW has the advantages over Earth-orbiting GRB monitors of continuous coverage, uninterrupted by Earth occultation, and a steady background, undistorted by passages through the Earth's trapped radiation, and subject only to occasional solar particle events. 

The instrument has two operational modes: waiting and triggered. 
In the waiting mode, the count rates are recorded in three energy windows: G1, G2, and G3, with 2.944~s time resolution. 
The nominal boundaries of the energy windows are 13--50~keV~(G1), 50--200~keV~(G2), and 200--750~keV~(G3).
When the count rate in the G2 window exceeds a $\sim 9 \sigma$ threshold above the background 
on one of two fixed timescales, 1~s or 140~ms, the instrument switches into the triggered mode. 
In the triggered mode, the count rates in G1, G2, and G3 are recorded with time resolutions varying from 2~ms up to 256~ms. 
Spectral measurements are carried out, starting from the trigger time, 
in two overlapping energy intervals, PHA1~(13--760~keV) and PHA2~(160~keV--10~MeV), 
with 64 spectra being recorded for each interval over a 63-channel, pseudo-logarithmic, energy scale.

The detector energy scale is calibrated in flight using the 1460~keV line of $^{40}$K and the 511~keV e$^+$ e$^-$ annihilation line in the multichannel spectra.
The gain of the detectors has slowly decreased during the long period of operation.
Accounting for the gain loss, the full KW spectral band in the waiting mode (G1+G2+G3) during 2005--2018 was $\sim$20--1500~keV for S1 and $\sim$17--1250~keV for S2;
the G1, G2, and G3 energy bounds shifted accordingly. A more detailed discussion of the KW instrument can be found in \citet{Svinkin2016} and T17.  

As of 2018~December, KW has triggered $\sim 4800$ times on a variety of transient events, including $\sim 3050$ GRBs.
Thus KW has been triggering on $\sim 130$ GRBs per year.
Additionally, the KW waiting mode data search \citep{Kozlova2019} revealed $\sim$2200 untriggered GRBs detected by KW and simultaneously by BATSE, BeppoSAX, \textit{Swift}, and \textit{Fermi} during the same time interval.
The KW databases of short bursts along with their IPN localization maps, GRBs with $z$, SGRs, and SFs can be found at the KW web page\footnote{\url{http://www.ioffe.ru/LEA/}}.

\subsection{\BAT}
The \textit{Swift} mission, dedicated to GRB studies, was launched on 2004~November~20~\citep{Gehrels2004}.
The \textit{Swift} payload is comprised of three instruments which work in tandem to provide rapid identification and multi-wavelength follow-up of GRBs and their afterglows: 
the Burst Alert Telescope (BAT), the X-ray Telescope (XRT; \citealt{Burrows2005}) and the UV-Optical Telescope (UVOT; \citealt{Roming2005}).
When the BAT detects a GRB, \textit{Swift} slews to the GRB position and observes the burst with the XRT and the UVOT, which can further refine the localization to $\lesssim$arcsec. 

The BAT is a highly sensitive, large field of view (FoV: 1.4~sr for $>50\%$ coded FoV and 2.2~sr for $>10\%$ coded FoV), coded-aperture telescope that detects and localizes GRBs in real time.
The BAT is composed of a detector plane that has 32,768 CdZnTe (CZT) detectors, and a coded-aperture mask that has $\sim$52,000 lead tiles.
The BAT energy range is 14--150~keV for imaging, which is a technique to subtract the background based on the modulation resulting from the coded mask; spectra with no position information can be obtained up to 350~keV.
Details of the BAT instruments, including the  in-orbit calibrations, can be found in \cite{Barthelmy2005} and the BAT GRB catalogs (\citealt{Sakamoto2008,Sakamoto2011,Lien2016}).

The sophisticated on-board localization capability of the BAT and the fast spacecraft pointing of \textit{Swift} allow X-ray (optical) afterglow observations from the XRT (UVOT) within a few hundred seconds after the trigger for more than 90\%~(30\%) of \textit{Swift} GRBs~\citep{Sakamoto2008}.
The precise and rapid GRB location provided by \textit{Swift} permits coordinated multi-wavelength observations on the ground that, in turn, provide a new opportunity to measure GRB redshifts and to use GRBs as cosmological probes.

\section{The Burst Sample}
\label{Sample}
The joint sample of the KW (waiting mode) and BAT bursts with known redshifts (hereafter, WM sample) was formed as follows.
First, we extracted a set of BAT GRBs with known redshifts from the online \textit{Swift} 
GRB Table\footnote{\url{https://swift.gsfc.nasa.gov/archive/grb_table/}} and excluded 
the bursts detected by KW in the triggered mode. 
Then we checked the reliability of the redshifts: we included only the bursts 
with $z$ measured from the emission lines, the absorption features of 
the host galaxies imposed on the afterglow continuum, or photometrically.

\begin{figure}
\center
\includegraphics[width=0.49\textwidth]{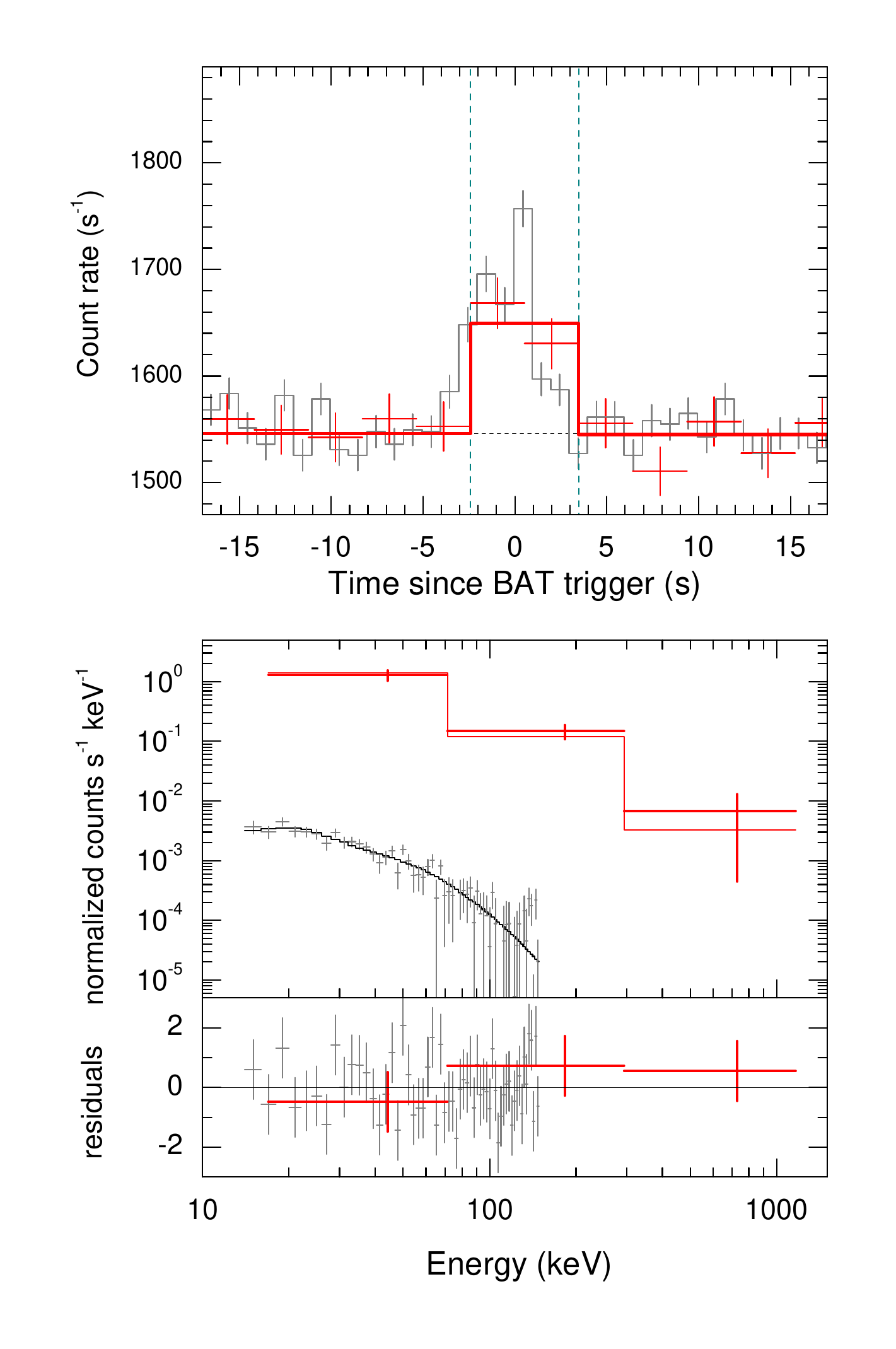}
\includegraphics[width=0.49\textwidth]{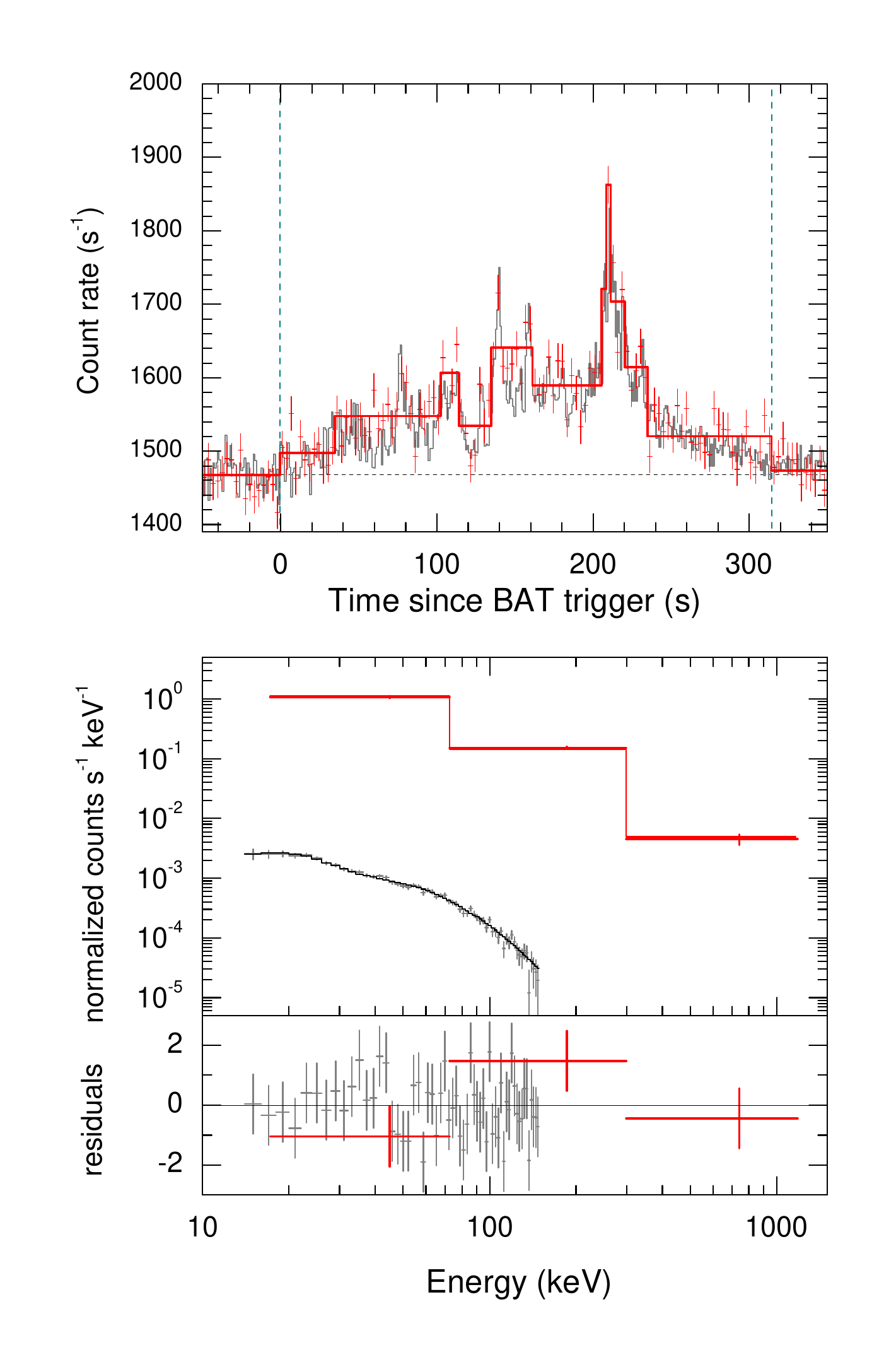}
\caption{Joint KW and BAT detections of GRB~090429B at $z$=9.38 (left) and GRB~110205A at $z$=2.22 (right). 
		The burst light curves are presented in the top panels. 
		The KW waiting-mode count rates (G1+G2, $\sim$20-350~keV) are shown with red points, and the horizontal dashed lines indicate KW background levels. 
		The BAT count rates (25-350~keV), arbitrary scaled to match the KW count rate and background, are shown with gray points and lines.
		Bayesian block divisions of the the KW light curves are plotted with thick red lines.
		The vertical dashed lines indicate time intervals chosen for time-integrated spectral fits.
		Joint KW+BAT fits to the TI spectra and the fit residuals are presented lower panels, where KW and BAT spectral points are shown by red and gray points, respectively. 
		The best-fit models (Table~\ref{spectraltab}) are plotted by solid lines: the Band function, with $\alpha\simeq-0.7$, $\beta\simeq-2.3$ and $E_\textrm{p}\simeq41$~keV, for GRB~090429B;
		and CPL ($\alpha\simeq-1.55$, $E_\textrm{p}\simeq220$~keV) for GRB~110205A.}
\label{LC_example}
\end{figure}

The remaining set comprised $\sim$300 bursts, for which, using a Bayesian block (BB) analysis \citep{Scargle2013}, 
we performed a targeted search for detections in the continuous KW waiting mode data. 
The decomposition of KW light curves in the G1+G2, G2, and G1+G2+G3 bands into BBs was performed on the 2.944~s time scale, using a custom tool\footnote{\url{https://github.com/dsvinkin/b_blocks}}.
The BB algorithm was tuned, for each energy band, to find intervals with $\gtrsim4$$\sigma$ count rate excess over background  (see, e.g. \citealt{Kozlova2019} for more details). 
The presence of such interval(s), temporally coincident with the BAT trigger\footnote{The KW data timing was corrected for the light propagation time between \textit{Wind} and \textit{Swift} for the specific burst.}, indicated a joint KW and BAT detection.
Using this criterion, we found 203 joint KW and BAT GRB detections 
and about a dozen events with lower KW signal significance ($\gtrsim3.2$$\sigma$), 
which were selected for the subsequent analysis manually.
Two examples of joint KW+BAT detections, GRB~090429B ($z$=9.38) and GRB~110205A ($z$=2.22), are shown in Figure~\ref{LC_example}.

Finally, we performed joint spectral fits to the 3-channel KW waiting mode spectra, 
covering the $\sim$20--1500~keV band, and the BAT spectra in the 15--150~keV band (see Section~\ref{Analysis} for details), 
and selected only the bursts appropriately fitted with a ``curved'' spectral model (CPL or BAND)
that allows $E_\textrm{p}$ to be constrained.
The final WM sample comprises 167 GRBs with reliable redshift estimates and constrained peak energies detected simultaneously by \BAT and by KW in the waiting mode from the beginning of the \textit{Swift} era in 2004 to the end of 2018.
Four bursts from this sample are short ones (GRB~050724, GRB~060801, GRB~090426, and GRB~131004A) and three bursts are X-ray flashes (XRFs: GRB~080515, GRB~091018, and GRB~120922A).
The general information about these bursts is presented in Table~\ref{generaltab}.
The first three columns contain the GRB name as reported in the Gamma-ray Burst Coordinates 
Network circulars\footnote{\url{http://gcn.gsfc.nasa.gov/gcn3\_archive.html}}, 
the BAT trigger ID and the BAT trigger time $T_0$. 
The next three columns provide the burst redshift and information on it.
For a number of GRBs there are several independent redshift estimates available, of which we gave a preference to spectroscopic over photometric redshift, if available; also, results from refereed papers, which presented a detailed spectral analysis, were given higher priority over earlier GCN circulars.
The rightmost two columns of Table~\ref{generaltab} contain the burst durations $T_{100}$ along with $t_0$, their start times relative to the BAT trigger time, determined using the BB decomposition of the BAT light curve in the 25--350 keV band.
This energy band was selected as, first, being common between both instruments and, 
second, being less sensitive to weak precursors or soft extended tails (T17). 
More details on the $T_{100}$ computation are given in Section~\ref{Spectral}.

\begin{figure}
\center
\includegraphics[width=0.8\textwidth]{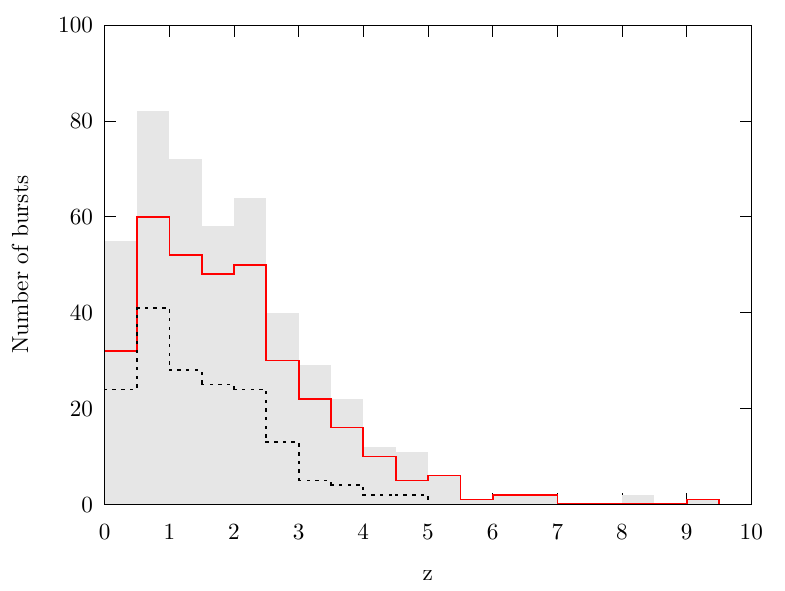}
\caption{GRB redshift distributions up to Jan~2019. Gray shaded area: all 469 GRBs with known redshifts (``GRBOX'').   
The red line: 338 KW GRBs (both triggered and waiting mode).
The black dotted line: 171 KW triggered bursts.} 
\label{redshiftdistr}
\end{figure}

The redshifts of GRBs in the WM sample span the range  $0.04 \leq z \leq 9.4$ and have mean and median values of $\sim 2.4$ and $\sim 2.2$, respectively.
These statistics are comparable with those for the \textit{Swift} era ($\bar{z}\sim 2.3$, \citealt{Coward2013}) 
but differ significantly from the statistics for the KW triggered bursts presented in T17, which were comparable with the pre-\textit{Swift} era ones \citep{Berger2005a}.
Figure~\ref{redshiftdistr} shows the redshift distributions for KW GRBs  
and all GRB redshifts available up to the end of 2018\footnote{Gamma-Ray Burst Online Index: \url{http://www.astro.caltech.edu/grbox/grbox.php}}.

\section{Data Analysis and Results}
\label{Analysis}
\subsection{Energy Spectra}
\label{Spectral}
We began the analysis with extraction of \BAT spectral data and light curves, performing the following steps.
First, we downloaded the data from the \BAT Enhanced GRB Data Product Catalog\footnote{\texttt{https://swift.gsfc.nasa.gov/results/batgrbcat/\textit{<GRB\_name>}/data\_product/\textit{<triger\_ID>}-results-detection-mask} (if it exists) or \texttt{https://swift.gsfc.nasa.gov/results/batgrbcat/\textit{<GRB\_name>}/data\_product/\textit{<triger\_ID>}-results}}. 
These folders contain the results from \citet{Lien2016} obtained with \textsc{batgrbproduct}, 
a standard automatic pipeline script for GRB processing, and some additional products required for the quick-look page.
The detailed description of the data processing procedure may be found in the abovementioned paper.
Then we constructed BAT light curves in the 25--350~keV range using a standard tool \textsc{batbinevt} from the software package \textsc{heasoft}.
After that we did a BB decomposition of the BAT light curve using the \textsc{battblocks} task.

Based on the BB decomposition of BAT (25--350~keV) and KW (G1+G2, $\sim$20--350~keV) light curves, we selected time intervals for spectral analysis.
For each burst in our sample, two time intervals were selected: time-averaged fits were performed over the interval closest to $T_{100}$ 
(hereafter the TI spectrum); the peak spectrum corresponds to the KW BB where the peak count rate (PCR) is reached. 
In some cases the $T_{100}$ and the spectral interval boundaries were adjusted after visual inspection.
The time interval between the beginning of the second BB and that of the last one was typically adopted as $T_{100}$.

Then we extracted BAT spectra for the selected intervals using the \textsc{batbinevt} tool and \textsc{batphasyserr} task to include systematics.
Since \textit{Swift} slews at a rate of $\sim 1^\circ$ per second \citep{Markwardt2007}, the telescope motion should be taken into account if the spectrum accumulation time overlaps with the spacecraft slew time by more than 5~s.
Thus, following \citet{Sakamoto2011} (but see also \citealt{Lien2016}), we created an ``average'' response file for the whole spectrum from multiple 5-s response files using the \textsc{HEASARC} tool \textsc{addrmf} with weighting factors equal to the fraction of light curve counts in the specific time periods of each response file.
The 5-s response files were made with the \textsc{batdrmgen} tasks, and the spectral files were updated using the \textsc{batupdatephakw} task.

We performed a joint spectral analysis of KW and BAT spectral data using \textsc{XSPEC} version 12.9.0 (\citealt{Arnaud1996}) using the $\chi^2$ statistic.
Each spectrum was fitted by two spectral models, both normalized to the energy flux in the observer-frame 15~keV--1.5~MeV range.
The first model is the Band function (hereafter BAND; \citealt{Band1993}):
\begin{equation}
\label{eq:Band}
f(E) \propto  \left\{ \begin{array}{ll}
E^{\alpha} \exp \left(-\frac{E (2+\alpha)}{E_\textrm{p}}\right), & \quad E<(\alpha-\beta)\frac{E_\textrm{p}}{2+\alpha} \\ E^{\beta} \left[(\alpha-\beta)\frac{E_\textrm{p}}{(2+\alpha)}\right]^{(\alpha-\beta)} \exp(\beta-\alpha), & \quad E
\ge(\alpha-\beta)\frac{E_\textrm{p}}{2+\alpha}, \\ \end{array} \right.
\end{equation}
where $\alpha$ is the low-energy photon index and $\beta$ is the high-energy photon index.
The second spectral model is an exponentially cutoff power-law (CPL), parameterized as $E_\textrm{p}$:
\begin{equation}
\label{eq:CPL}
f(E) \propto E^{\alpha} \exp \left(-\frac{E (2+\alpha)}{E_\textrm{p}}\right).
\end{equation}

The fits were performed in the energy range from $\sim 15$~keV to $\sim 1.5$~MeV.
When fitting the BAT spectral data, channels 1-3 ($\lesssim 14$~keV) and the channels starting from 63 ($\gtrsim 149$~keV) were ignored.
The parameter errors were estimated using the \textsc{XSPEC} command \textsc{error} based on the change in fit statistic ($\Delta \chi^2 = 1$) which corresponds to the 68\% confidence level (CL).
The models were multiplied by a constant normalization factor to take into account the systematic effective area uncertainties in the response matrices of each instrument.
The normalization factor of the KW data was fixed to unity, and the normalization factor of the BAT spectra was the free parameter. 
The latter, for our sample, is distributed approximately normally with mean $\approx$0.80 and $\sigma \approx 0.11$, in general agreement with the conclusion of \citet{Sakamoto2011a} that the intrinsic effective area of the BAT energy response file is 10\%--20\% smaller than that of the KW.

Examples of joint KW+BAT spectral fits for TI spectra of GRB~090429B and GRB~110205A are given in Figure~\ref{LC_example}.
The results of the spectral analysis of the WM sample are presented in Table~\ref{spectraltab}.
For each spectrum, we present the results for the models whose parameters are constrained (hereafter, GOOD models).
The spectral fits with the BAND model with $\beta < -3.5$, which are typically poorly constrained, were excluded.
The ten columns in Table~\ref{spectraltab} contain the following information: (1) the GRB name (see Table~\ref{generaltab}); (2) the spectrum type, where ``i'' indicates that the spectrum is time-integrated (TI), ``p'' means that the spectrum is peak (for some bursts with poor count statistics, the TI and the peak spectra are measured over the same interval); (3) and (4) contain the spectrum start time $t_\textrm{start}$ (relative to $T_0$) and its accumulation time $\Delta T$; (5) GOOD models for each spectrum; (6)--(8) $\alpha$, $\beta$, and $E_\textrm{p}$; (9) $F$ (normalization flux); (10) $\chi ^{2}/\textrm{d.o.f.}$ along with the null hypothesis probability given in brackets.
In cases where the lower limit for $\beta$ is not constrained, the value of ($\beta_\textrm{min} - \beta$) is provided instead, where $\beta_\textrm{min} = -10$ is the lower limit for the fits.

We found BAND to be a GOOD model for 56 TI and 50 peak spectra. The remaining spectra were well-fitted only by the CPL function.
Table~\ref{stattab} summarizes the descriptive statistics for spectral parameters and energetics for 
the WM sample from this paper and the whole sample of KW GRBs with known redshifts.
$E_\textrm{p}$ varies from $\approx 27$~keV to $\approx 580$~keV for the WM sample and lies within a significantly wider range, from $\approx 27$~keV to $\approx 3.7$~MeV, for the whole KW sample.
The TI spectrum $E_\textrm{p}$ ($E_\textrm{p,i}$) has its median value at 111~keV, while the peak spectrum $E_\textrm{p}$ ($E_\textrm{p,p}$) median value is 136~keV.
The rest-frame peak energies of the whole KW sample, $E_\textrm{p,i,z} = (1+z) E_\textrm{p,i}$ and $E_\textrm{p,p,z} = (1+z) E_\textrm{p,p}$, cover a wide range from $\approx 53$~keV  to $\approx 7$~MeV (GRB~090510).

\subsection{Burst Energetics}
\label{Energetics}
The energy fluences ($S$) and the peak energy fluxes ($F_\textrm{peak}$) were derived using the 15~keV--1.5~MeV normalization fluxes 
from the spectral fits with the BAND model (or the CPL function if the fit with the BAND model was not constrained) for TI and peak spectra, respectively.
Since the TI spectrum accumulation interval may differ from the $T_{100}$ interval, a correction which accounts for the emission outside the TI spectrum was introduced when calculating $S$.
Following T17 three time scales $\Delta T_\textrm{peak}$ were used when calculating $F_\textrm{peak}$: 1~s, 64~ms, and the ``rest-frame 64~ms'' scale ($(1 + z) \cdot 64$~ms); the latter was used to estimate the rest-frame peak luminosity $L_\textrm{iso}$.
To obtain $F_\textrm{peak}$, the model energy flux of the peak spectrum was multiplied by the ratio of the PCR on the $\Delta T_\textrm{peak}$ scale to the average count rate in the spectral accumulation interval. Both corrections were calculated using counts in the BAT light curve in the 25--350~keV band.

The cosmological rest-frame energetics, the isotropic-equivalent energy release $E_\textrm{iso}$ and the isotropic-equivalent peak luminosity $L_\textrm{iso}$, can be calculated as $E_\textrm{iso} = \frac{4 \pi D_\textrm{L}^2}{1+z} \times S \times k$ and $L_\textrm{iso} = 4 \pi D_\textrm{L}^2 \times F_\textrm{peak} \times k$, with the proper $k$-correction, which transforms the energetics from the observer-frame  1.5~keV--1.5~MeV energy range to the $1/(1 + z)$~keV--10~MeV band.

Knowing $t_{\mathrm{jet}}$, one can estimate the collimation-corrected energy released in gamma-rays $E_{\gamma} = E_{\textrm{iso}} (1-\cos \theta_\textrm{jet})$ and the collimation-corrected peak luminosity $L_{\gamma} = L_{\textrm{iso}} (1-\cos \theta_\textrm{jet})$, where $\theta_\textrm{jet}$ is the jet opening angle and $(1-\cos \theta_\textrm{jet})$ is the collimation factor.
We consider two types of circumburst medium (CBM): a CBM with constant number density $n$, hereafter homogeneous medium, or HM, and a stellar-wind-like CBM with $n(r) \propto r^{-2}$, hereafter SWM.
In this work, we only use the jet breaks that were detected either in optical/IR afterglow light curves or in two spectral bands simultaneously.
We found reasonably-constrained jet break times for 14 bursts in the literature (including one short GRB), and for six of them the HM was taken to be the most probable CBM.
See T17 for further details on the calculation of jet opening angles.

Table~\ref{energytab} summarizes observer-frame and isotropic-equivalent rest-frame energetics. The first two columns are the GRB name and $z$.
The next seven columns present the observer-frame energetics: $S$; peak fluxes on the three time scales: $F_\textrm{peak,1000}$ (1~s), $F_\textrm{peak,64}$ (64~ms), and $F_\textrm{peak,64,r}$ ($(1 + z) \cdot 64$~ms); together with the start times of the intervals when the PCR is reached: $T_\textrm{peak,1000}$, $T_\textrm{peak,64}$, and $T_\textrm{peak,64,r}$. 
The rightmost two columns contain $E_\textrm{iso}$ and the peak isotropic luminosity, $L_\textrm{iso}$, calculated from $F_\textrm{peak,64,r}$.
These $L_\textrm{iso}$ values may be adjusted to a different time scale $\Delta T$ (64~ms or 1~s) as: 
$$L_\textrm{iso}(\Delta T)=\frac{F_\textrm{peak}(\Delta T)}{F_\textrm{peak,64,r}} L_\textrm{iso}.$$
The most fluent burst in our sample is GRB~110205A ($S \approx 4.1 \times 10^{-5}$~erg~cm$^{-2}$).
The brightest burst based on the peak energy flux is GRB~050724 ($F_\textrm{peak,64,r} \approx 2.9 \times 10^{-6}$~erg~cm$^{-2}$~s$^{-1}$).
The most energetic burst in terms of the isotropic energy is GRB~050904 ($E_\textrm{iso} \approx 1.4 \times 10^{54}$~erg).
The most luminous burst is GRB~130606A ($L_\textrm{iso} = 3.6 \times 10^{53}$~erg~s$^{-1}$).

The collimation-corrected energetics for 14 bursts with ``reliable'' jet break times is presented in Table~\ref{collimationtab}.
The first column is the burst name. 
The next three columns specify $t_\textrm{jet}$, the CBM environment implied, and references to them.
The next columns contain the derived jet opening angles and the corresponding collimation factors, and the last two columns present $E_\gamma$ and $L_\gamma$.
For bursts with no reasonable constraint on the CBM profile the results are given for both HM and SWM.
The HM jet opening angles vary from $1.3\degree$ to $10.2\degree$ and the corresponding collimation factors from $0.0002$ to $0.016$.
The brightest burst in terms of both $E_{\gamma}$ and $L_{\gamma}$ is GRB~060418 ($E_{\gamma,\textrm{HM}} \simeq 2.3 \times 10^{51}$~erg, $L_{\gamma,\textrm{HM}} \simeq 2.9 \times 10^{50}$~erg~s$^{-1}$, $\theta_\textrm{jet,HM} \simeq 10.0\degree$).

\subsection{Comparison with  T17 Results}
\label{Discussion_comparison}
\begin{figure}
\center
\includegraphics[width=0.8\textwidth]{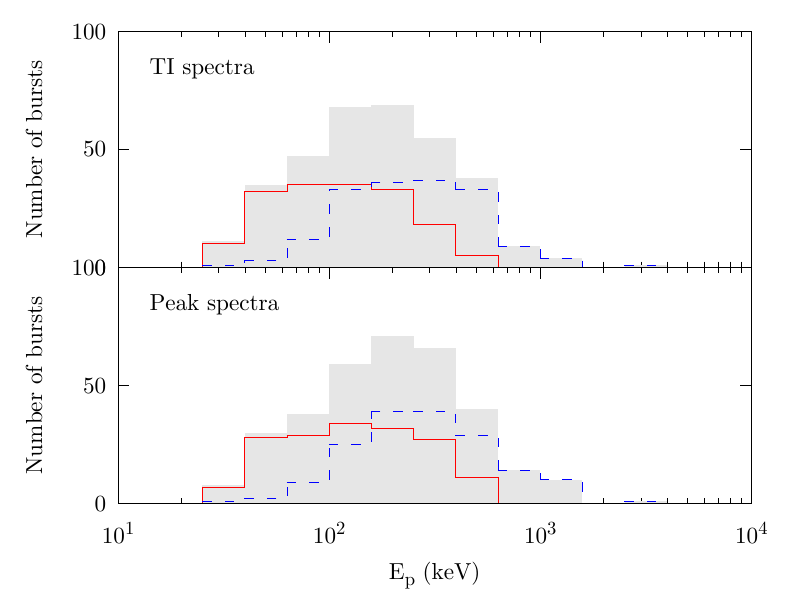}
\caption{Distributions of $E_\textrm{p}$ (observer frame) for the WM sample (red solid lines), the KW triggered sample (blue dashed lines), 
and the whole KW sample of GRBs with known redshifts (the gray shaded areas).}
\label{GraphEp}
\end{figure}

Concluding the WM sample analysis, we note some differences between the results obtained in this work and the ones for the sample of triggered KW bursts (T17).
The first is a bias towards fainter and spectrally-softer GRBs in this catalog that comes from different detection algorithms.
The KW hardware trigger operates at 9$\sigma$ threshold in the $\sim$80--350~keV band and strongly suppresses the detection of bursts with $E_\textrm{p}\lesssim100$~keV (Section~5.3 in T17). Accordingly, the T17 sample is comprised of relatively bright and spectrally-hard GRBs, with the median $E_\textrm{p}$ of $\sim240$~keV and $F_\textrm{peak}\gtrsim 1 \times 10^{-6}$~erg~cm$^{-2}$~s$^{-1}$.
The KW WM sample, that excludes the trigged events, was formed through a targeted search of BAT GRBs in the continuous KW waiting-mode data (Section~\ref{Sample}). 
The search was performed in several KW energy bands, starting from $\sim$20~keV, with an effective detection threshold in each band of $\sim4$$\sigma$.
As a result, the median $E_\textrm{p}$ in the WM sample ($\sim115$~keV) is a factor of two lower than that of T17 (Figure~\ref{GraphEp}); 
the brightest WM burst has $F_\textrm{peak}$ just a few times higher than the T17 lower limit, $\sim 3 \times 10^{-6}$~erg~cm$^{-2}$~s$^{-1}$,
while the weakest one has $F_\textrm{peak}\sim 1 \times 10^{-7}$~erg~cm$^{-2}$~s$^{-1}$.
Thus, thanks to the high sensitivity of BAT, the sample-wide flux threshold has been reduced by an order of magnitude as compared to T17. 

In T17, burst energetics were estimated using the flux of the BEST spectral model chosen based on the difference in $\chi^2$ between 
the CPL and the BAND fits to KW multichannel spectra. 
The criterion for accepting the model with the single additional parameter was the change in $\chi^2$ of at least 6 ($\Delta \chi^2 \equiv \chi_\textrm{CPL}^2 - \chi_\textrm{BAND}^2 > 6$).
For the WM sample, the high-energy part of the analyzed spectrum (above 150~keV) is represented by just two wide KW waiting mode spectral channels, 
which, in many cases, allowed us to constrain the high-energy photon index $\beta$ (and, simultaneously, better constrain $E_p$ as compared to a CPL fit to the same spectrum). 
On the other hand, with the lack of detailed spectral coverage at the higher energies, our fits with the two models 
did not typically result in a statistically-significant $\Delta \chi^2$ between the BAND and CPL fits. 
Thus, in order not to underestimate the high-energy flux we calculated the burst energetics using the BAND model fit when it is constrained.

\cite{Atteia2017} and T17 found a good agreement between KW- and \textit{Fermi}/GBM-derived energetics 
and spectra of several dozen commonly detected bright GRBs with known redshifts. 
The WM sample studied in this work contains 45 faint GRBs independently detected by \textit{Fermi}/GBM. 
For $\sim$30 GBM bursts with constrained peak energies, we compared the $E_\textrm{p}$ of time-integrated spectra and fluences reported in the GBM catalog\footnote{\url{https://heasarc.gsfc.nasa.gov/W3Browse/fermi/fermigbrst.html}} with those obtained in this work. 
In 27 cases where the KW+BAT and the GBM fits were made over comparable time intervals, we found them to agree within errors.

\section{Discussion}
\label{Discussion}
\subsection{The progress achieved with this catalog}
This work extends the first part of the KW catalog of GRBs with known redshifts (T17) 
which provided temporal and spectral parameters, as well as broadband observer- and rest-frame energetics of 150 bursts 
detected in the KW triggered mode in a wide $\sim$20~keV--20~MeV energy range, with the sample-wide 
peak flux limit of $\sim 1 \times 10^{-6}$~erg~cm$^{-2}$~s$^{-1}$. Extended up to the end of 2018, this sample comprises 171 GRBs with $0.1 \leq z \leq 5$.

With 167 GRBs from the WM sample added, the total number of KW bursts with known redshifts increases to 338 GRBs, 
the largest set of cosmological GRBs studied to date over a broad energy band.
It includes 43 long bursts with reliable $t_\mathrm{jet}$ estimates, and 14 short/hard ``type~I'' GRBs.  
With the flux limit of $F_\textrm{peak}\sim 1 \times 10^{-7}$~erg~cm$^{-2}$~s$^{-1}$ (Figure~\ref{GraphGRBFR}), the KW sample spans redshifts $0.04 \leq z \leq 9.4$, 
isotropic luminosities $L_\mathrm{iso}$ from $\sim 2 \times 10^{48}$~erg~s$^{-1}$ to $\sim 5 \times 10^{54}$~erg~s$^{-1}$, 
isotropic energies $E_\mathrm{iso}$ from $\sim 3 \times 10^{49}$~erg to $\sim 6 \times 10^{54}$~erg, and intrinsic peak energies from $\sim$50~keV to $\sim$7~MeV.
At this time, we do not include six KW ultra-long (with durations $>$1000~s) GRBs with known redshifts in this sample;
their properties will be reported elsewhere.

\subsection{Hardness-Intensity Correlations}
\label{Correlations}
The large set of measured GRB redshifts, together with well-determined prompt emission spectra and fluences, 
can provide an excellent testing ground for the widely discussed correlations between rest-frame spectral hardness 
and energetics, e.g., the “Amati” \citep{Amati2002}, “Yonetoku” \citep{Yonetoku2004} or “Ghirlanda” \citep{Ghirlanda2007} relations. 
This could facilitate using GRBs as standard candles (see, e.g., \citealt{Atteia1997} or Friedman \& Bloom 2005) 
and probing cosmological parameters with GRBs (see, e.g., Cohen \& Piran 1997 or Diaferio et al. 2011).

Using the updated KW sample, and following the methodology of T17, we tested the rest-frame Amati ($E_\textrm{p,i,z}$--$E_\textrm{iso}$) and Yonetoku ($E_\textrm{p,p,z}$--$L_\textrm{iso}$) correlations, 
along with their collimated versions, $E_\textrm{p,i,z}$--$E_\gamma$ and $E_\textrm{p,p,z}$--$L_\gamma$.
To probe the existence of correlations, we calculated the Spearman rank-order correlation coefficients ($\rho_S$) and the associated null-hypothesis (chance) probabilities or p-values ($P_{\rho_S}$; \citealt{Press1992}). 
We approximated linear regression between $\textrm{log}$-energy and $\textrm{log}-E_\textrm{p}$ using two methods, with and without intrinsic scatter ($\sigma_\textrm{int}$).
The fit was performed using the \textsc{MPFITEXY} routine\footnote{\url{http://purl.org/mike/mpfitexy}} \citep{Williams2010}, which depends on the \textsc{MPFIT} package \citep{Markwardt2009}.

\begin{figure}
\center
\includegraphics[width=0.49\textwidth]{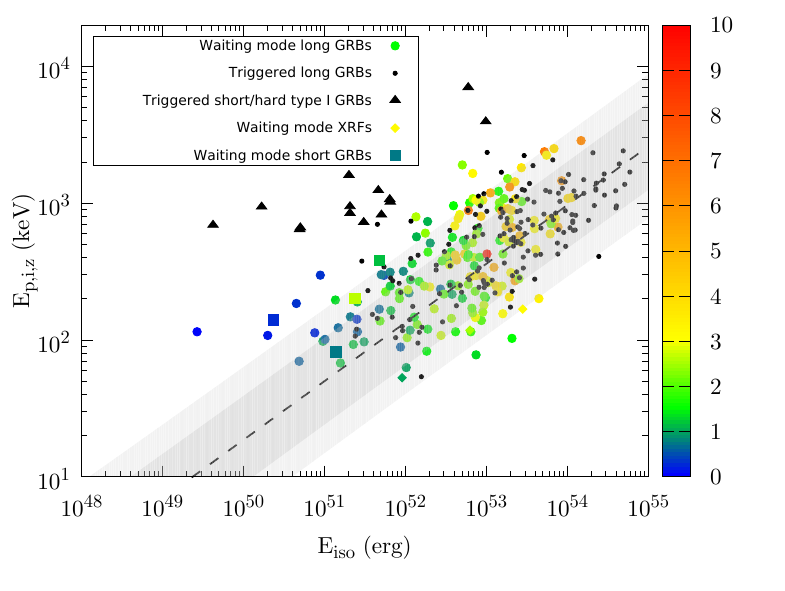}
\includegraphics[width=0.49\textwidth]{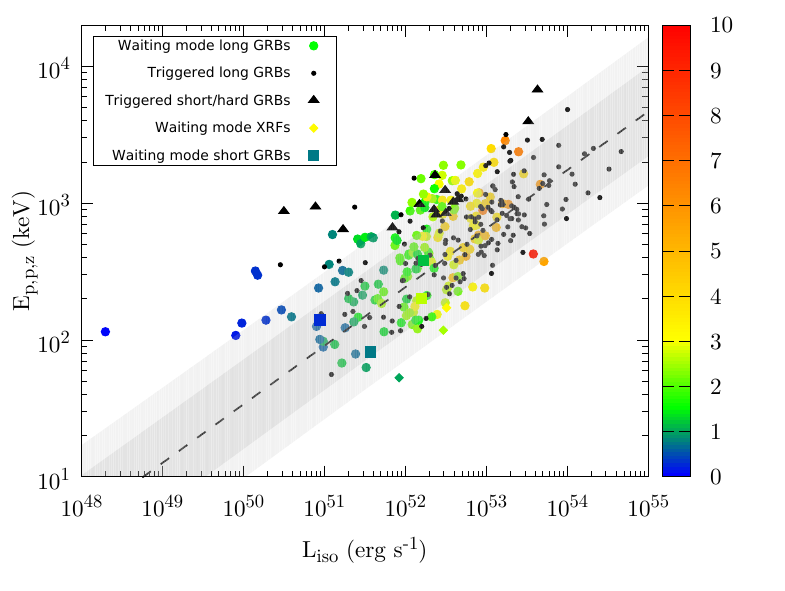}
\caption{
Rest-frame energetics of 338 KW GRBs in the $E_\textrm{iso}–-E_\textrm{p,i,z}$ (left) and $L_\textrm{iso}–-E_\textrm{p,p,z}$ (right) planes.
Black symbols: KW triggered bursts.
Colored symbols: the WM sample (this work); the color of each data point represents the GRB redshift.
The ``Amati'' and ``Yonetoku'' relations calculated for 315 KW long (type~II) GRBs are plotted with dashed lines, and
the dark- and light-gray shaded areas show their 68\% and 90\% PI's, respectively.
} 
\label{AmatiYonetoku}
\end{figure}

The correlation parameters obtained for the subsample of 316 long (or type~II) GRBs, and 43 long bursts with $t_\textrm{jet}$ estimates are summarized in Table~\ref{correlationtab}.
The first column presents the correlation. The next three columns provide the number of bursts in the fit sample, $\rho_S$, and $P_{\rho_S}$.
The next columns specify the slopes ($a$), the intercepts ($b$), and $\sigma_\textrm{int}$.
The derived slopes of the Amati and Yonetoku relations are very close to each other, 0.429 ($\rho_S$=0.70, 316 GRBs) and 0.428 ($\rho_S$=0.70, 316 GRBs), respectively.
When accounting for the intrinsic scatter, these slopes change to a more gentle $\sim$0.31 (with $\sigma_\textrm{int}\sim$0.23).
It should be noted that the slopes of both correlations for the extended GRB sample are slightly shallower than those for the sample from T17.
Figure~\ref{AmatiYonetoku} represents the correlations for the long (type~II) GRBs from the KW sample.

The subsample of GRBs with reliable $t_\textrm{jet}$ estimates comprises 43 long bursts.
The Ghirlanda relations were tested against the Amati and Yonetoku relations for the same subset of GRBs.
The geometric mean of the HM and SWM collimation factors was used for the bursts with unknown CBM.
With the whole KW sample we confirm the result of T17 that accounting for the jet collimation neither improves the significance of the correlations nor reduces the dispersion of the points around the best-fit relations.
The slopes we obtained for the collimated Amati and Yonetoku relations are steeper compared to those of the non-collimated versions.
\subsection{GRB Luminosity and Isotropic-energy Functions, and GRB Formation Rate}
\label{GRBFR}
With the updated sample of 315 long KW GRBs with known $z$, we estimated the luminosity function (LF; the number of bursts per unit luminosity), the isotropic energy release function (EF; the number of bursts per unit energy release), and the cosmic GRB formation rate (GRBFR; the number of events per comoving volume and time). 
Our approach (details are given in T17) relies on the non-parametric Lynden-Bell C$^-$ technique \citep{Lynden-Bell1971} further advanced by \cite{Efron1992} (LBEP method). 

Without loss of generality, the total LF $\Phi(L_\textrm{iso},z)$\footnote{Similar reasoning may be applied to the total EF $\Psi(E_\textrm{iso},z)$} can be rewritten as $\Phi(L_\textrm{iso},z) = \rho(z)\phi(L_\textrm{iso}/g(z), \mathbf{\zeta})/g(z),$
where $\rho(z)$ is the GRB formation rate (GRBFR);
$\phi(L_\textrm{iso}/g(z))$ is the local LF; $g(z)$ is the luminosity evolution that parameterizes the correlation between $L$ and $z$;
and $\mathbf{\zeta}$ stands for the cosmological evolution of the LF shape parameters, whose effect is commonly ignored as the shape of the LF does not change significantly with $z$ (e.g. \citealt{Yonetoku2004}).

Following \citet{Lloyd-Ronning2002}, \citet{Yonetoku2004}, \citet{Wu2012}, and \citet{Yu2015} we chose the functional form $g(z) = (1 + z)^\delta$ for the luminosity evolution.
It should be noted that the isotropic luminosity evolution can be determined by either the evolution of the amount of energy per unit time emitted by the GRB progenitor or by the jet opening angle evolution (see, e.g., \citet{Lloyd-Ronning2002} for the discussion); we tested the KW sample for a correlation between the collimation factor and $1 + z$ and, as in T17, found the correlation negligible (the Spearman rank-order correlation coefficient $\rho_S \approx -0.4$, and the corresponding $p$-value $P_{\rho_S} \approx 0.01$) for the subsample of 43 long (or type~II) bursts with known collimation factors.
See T17 (Section~5.5 and Appendix) for further details on the non-parametric method used for the calculations.

\begin{figure}
\center
\includegraphics[width=0.49\textwidth]{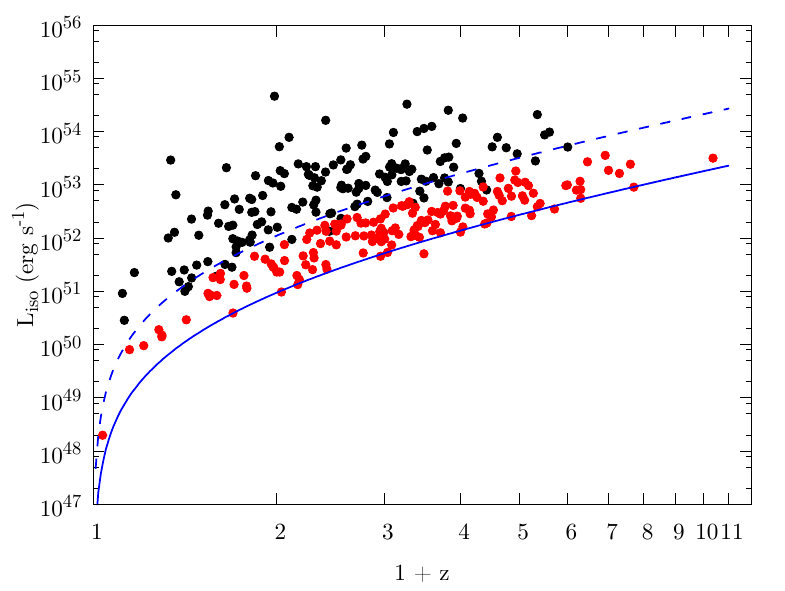}
\includegraphics[width=0.49\textwidth]{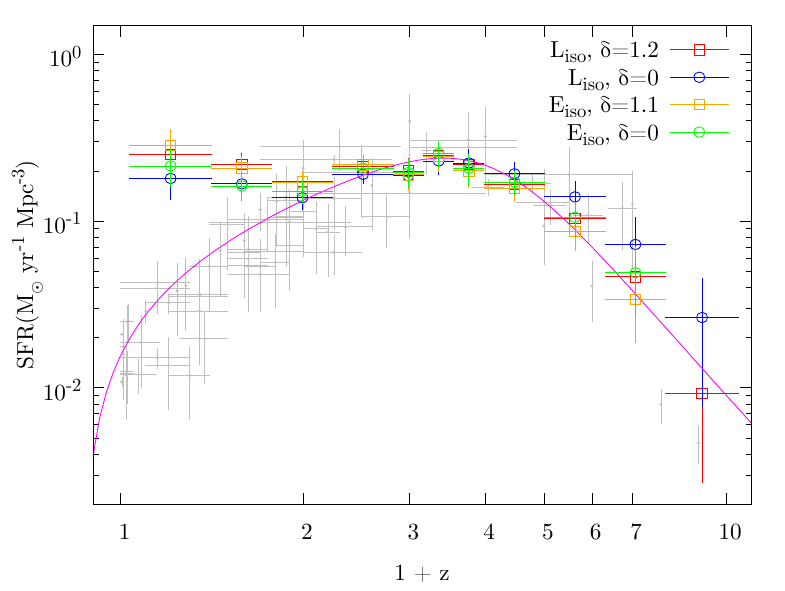}
\makebox[0.49\textwidth]{(A)}
\makebox[0.49\textwidth]{(B)}
\includegraphics[width=0.49\textwidth]{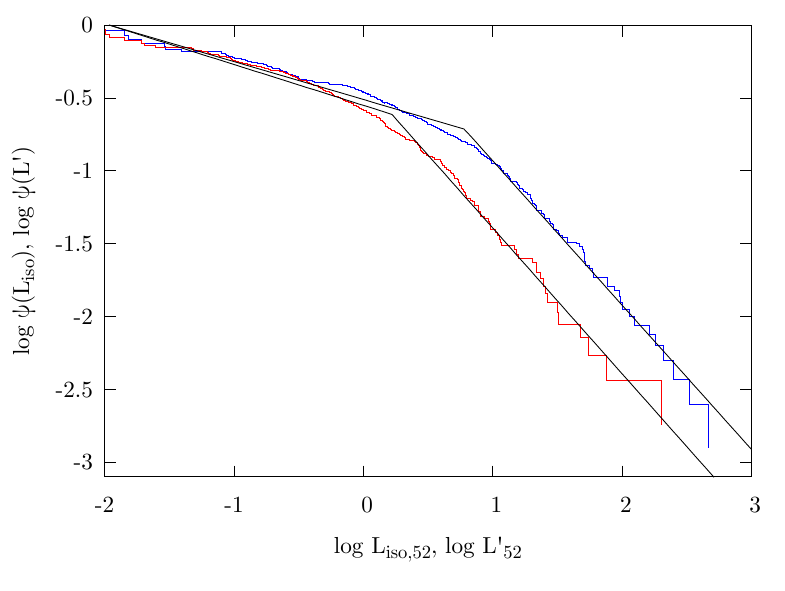}
\includegraphics[width=0.49\textwidth]{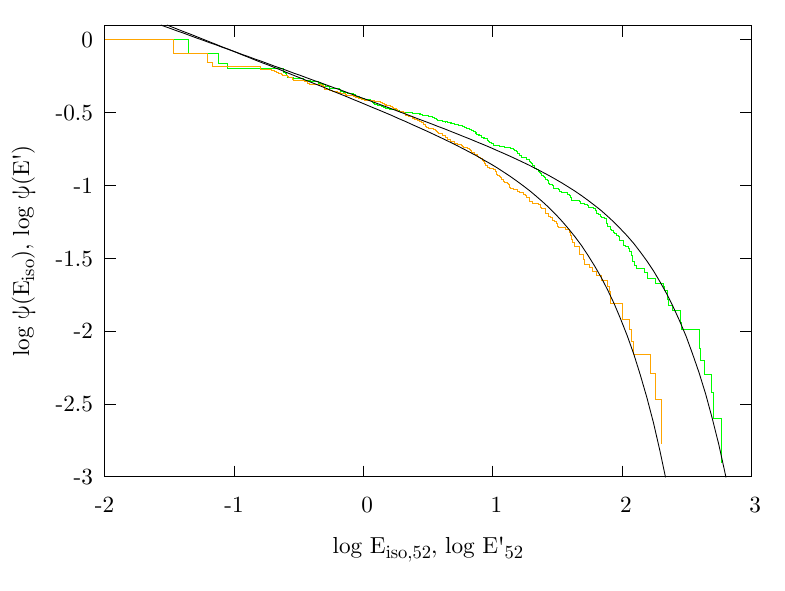}
\makebox[0.49\textwidth]{(C)}
\makebox[0.49\textwidth]{(D)}
\caption{
Panel~A: $L_\textrm{iso}$ vs. redshift for 315 KW GRBs.
Black symbols: 171 triggered bursts (the updated T17 sample).
Red symbols: the WM sample (167 bursts, this work).
The observer-frame flux limits $F_\mathrm{peak}$ (Section~\ref{GRBFR}) are shown by a dashed line (triggered bursts) and a solid line (the full KW sample).
Panel~B: Comparison of the derived GRBFR and the SFR data from the literature.
The gray points show the SFR data from \citet{Hopkins2004}, \citet{Bouwens2011}, \citet{Hanish2006}, and \citet{Thompson2006}.
The solid line denotes the SFR approximation from \citet{Li2008}.
The GRBFR normalization is the same for the results derived using the four datasets and the GRBFR points have been shifted arbitrarily to match the SFR at $(1+z)\sim3.5$.
Panel C shows cumulative GRB isotropic-luminosity functions (LFs: $\psi(L_\textrm{iso})$ - blue, $\psi(L')$ - red) and their BPL approximations (black). 
	Panel D shows cumulative GRB isotropic-energy functions (EFs: $\psi(E_\textrm{iso})$ - green, $\psi(E')$ - orange) and their exponentially cutoff PL approximations (black). 
	The distributions (Section~\ref*{GRBFR}) are normalized to unity at the dimmest points. The approximation parameters are given in Table~\ref{tabLFEF}.
} 
\label{GraphGRBFR}
\end{figure}

Since the LBEP method was specifically designed to reconstruct the intrinsic distributions from the observed distributions, takes into account the data truncations introduced by observational bias, and includes the effects of the possible correlation between the two variables, the correct estimation of the threshold fluxes and fluences plays a crucial role in the application of this technique.
The first panel of Figure~\ref{GraphGRBFR} shows the distribution of  KW bursts in the $L_\textrm{iso}$--$z$ diagram.
The blue dashed line corresponds to the truncation flux $F_\textrm{lim} \sim 2.0 \times 10^{-6}$~erg~cm$^{-2}$~s$^{-1}$ for the triggered sample of type~II GRBs from T17, while the solid line represents the truncation flux $F_\textrm{lim} \sim 1.7 \times 10^{-7}$~erg~cm$^{-2}$~s$^{-1}$ for the joint sample of waiting mode and triggered KW GRBs. The threshold fluence used for the EF estimates is $S_\textrm{lim} \sim 1.6 \times 10^{-6}$~erg~cm$^{-2}$.

Applying the LBEP method to the $z$-$L_\textrm{iso}$ plane for the long (type~II) bursts from the joint sample of KW GRBs with z, we found that the independence of the variables is rejected at $\sim 1.8 \sigma$, and the best luminosity evolution index is $\delta_\textrm{L} = 1.2_{-0.6}^{+0.6}$.
Applying the same method to the $z$-$E_\textrm{iso}$ plane of the same sample we found that the independence of the variables is rejected at $\sim 1.8 \sigma$, and the best isotropic energy evolution index is $\delta_\textrm{E} = 1.1_{-0.6}^{+1.0}$.
Note that the estimated $E_\textrm{iso}$ and $L_\textrm{iso}$ evolutions are comparable.
The evolution PL indices $\delta_L$ and $\delta_E$ derived here are shallower than those reported in the previous studies: $\delta_L=2.60^{+0.15}_{-0.20}$ (\citealt{Yonetoku2004}), $\delta_L=2.30^{+0.56}_{-0.51}$ (\citealt{Wu2012}), $\delta_L=2.43^{+0.41}_{-0.38}$ (\citealt{Yu2015}), and $\delta_E=1.80^{+0.36}_{-0.63}$ (\citealt{Wu2012}), and comparable with the indices from T17 ($\delta_L=1.7^{+0.9}_{-0.9}$, $\delta_E=1.1^{+1.5}_{-0.7}$).

The luminosity and energy release evolution can be eliminated dividing by $g(z)$: $L'=L_\textrm{iso}/(1+z)^{\delta_L}$ and $E'=E_\textrm{iso}/(1+z)^{\delta_E}$, where $L'$ and $E'$ are the local luminosity and energy release, correspondingly.
Then the cumulative distributions\footnote{$\phi(x) = -d\psi(x)/dx$} $\psi (L)$, $\psi (L')$, $\psi (E)$, and $\psi (E')$ were fitted
with a broken power-law (BPL) function:
$$ \psi(x) \propto \left\{ \begin{array}{ll}
x^{\alpha_1}, & \quad x \leq x_b, \\
x_b^{(\alpha_1 - \alpha_2)} x^{\alpha_2}, & \quad x > x_b, \\
\end{array} \right. $$
where $\alpha_1$ and $\alpha_2$ are the PL indices of the dim and bright distribution segments, and $x_b$ is the breakpoint of the distribution;
and with the CPL function\footnote{The CPL function definition is different here from that in Section~\ref{Spectral}}: $\psi(x) \propto x^\alpha\:\textrm{exp}(-x/x_\textrm{cut})$,
where $x_\textrm{cut}$ is the cutoff luminosity (or energy).

The fits were performed in $\textrm{log}-\textrm{log}$ space; the results are given in Table~\ref{tabLFEF} and shown in Figure~\ref{GraphGRBFR} (bottom panels).
In Table~\ref{tabLFEF} the central values of parameters are the best-fit values derived for the original data sets, 
and the upper and lower uncertainties are calculated from the 68\% quantiles computed for 10000 bootstrapped samples.
The derived BPL slopes of the LF and EF are close to each other, both for the dim and bright segments, so the shape of the EF is similar to that of the LF.
The BPL indices derived for the LF are in agreement with those obtained in \citet{Yonetoku2004} ($\alpha_1 = -0.29 \pm 0.02$ and $\alpha_2 = -1.02 \pm 0.02$), while the indices computed in \citet{Yu2015} are shallower ($\alpha_1 = -0.14 \pm 0.02$ and $\alpha_2 = -0.70 \pm 0.03$); the BPL indices derived for the EF are marginally consistent with those obtained in \citet{Wu2012} ($\alpha_1 = -0.27 \pm 0.01$ and $\alpha_2 = -0.87 \pm 0.07$).
Using the bootstrap approach we found that BPL better fits the $\psi (L)$ for $\sim 95$\% of the datasets, while CPL better fits the $\psi (E)$ for $\sim 70$\% of bootstrapped sets.
The existence of a sharp cutoff of the isotropic energy distribution of KW and \emph{Fermi}/GBM GRBs around $\sim 1-–3 \times 10^{54}$~erg was suggested by \cite{Atteia2017} and confirmed by T17.

We estimated the cumulative GRB number distribution $\psi (z)$ and the derived GRBFR per unit time per unit comoving volume $\rho(z)$ using the LBEP method.
In Panel B of Figure~\ref{GraphGRBFR} we compare the star formation rate (SFR) data from the literature  with GRBFRs derived from different $z$-$L$ and $z$-$E$ distributions.
The GRBFR estimated from all distributions (both corrected and uncorrected for the cosmological evolution of energetics) shows a relative excess over the SFR at $z<1$ and nearly traces the SFR at higher redshifts.
The low-$z$ GRBFR excess over SFR was reported in \citet{Yu2015}, \citet{Petrosian2015} and T17; this work confirms those findings at a higher significance. 

We performed simulations to test whether this relative excess of the GRBFR over the SFR is an artefact of the LBEP method.
First, we generated a set of 300 $(z,L_\mathrm{iso})$ pairs, with $L_\mathrm{iso}\gtrsim10^{47}$~erg~s$^{-1}$ that follow the cumulative distribution $\psi (L)$ estimated above; 
a redshift distribution following the SFR approximation \citep{Li2008}; and the sample truncating flux $F_\textrm{lim} = 1.7 \times 10^{-7}$~erg~cm$^{-2}$~s$^{-1}$, similar to that for the KW sample.
Then, the cosmological evolution $g(z)$ with $\delta_L=1.2$ (as estimated above) was applied to the simulated luminosities and the final set of $(z,L')$ pairs was generated from $(z,L)$.
We processed the simulated $(z,L')$ sample using the LBEF methodology described above and obtained a $\rho(z)$ that perfectly traces the seed SFR in the whole range of simulated redshifts $0.01 \lesssim z \lesssim 10$. 
The simulations were repeated several times with similar results. Thus, the implied low-$z$ GRBFR excess cannot be attributed to our analysis and may be intrinsic.

One of the possible keys to explain this relative excess of GRBFR over SFR may be the preference of long GRBs for low-mass galaxies and low-metallicity environments which are not unbiased tracers of the star formation rate at low redshifts (see, e.g., \citealt{Lloyd-Ronning2019} and references therein for further discussion).
Another reason may be the presence of selection effects based on the incompleteness of the sample: 
it is easier to measure the redshifts of nearest GRBs thereby creating a bias towards them which, in turn, will lead to the relative excess of low-$z$ 
bursts over the rest of the unbiased sample (see \citealt{Pescalli2016}).
The ``external'' selection effects (i.e. not related to the prompt emission), arising from the complex procedure of redshift measurements, 
are very important but extremely hard to estimate (see, however, \cite{Petrosian2015} who made an attempt to estimate the selection effect in X-rays). 
At least three biases may contribute to this: the ability to localize the GRB precisely, 
the ability of optical/NIR telescopes to start following the afterglow rapidly, 
and the factors affecting the detectability of spectral lines in the afterglow spectra.
Although the ``external'' biases are very important, their investigation lies beyond the scope of this paper.

\section{Summary and Conclusions}
\label{Conclusions}
We have presented the results of a systematic study of GRBs with reliable redshift estimates simultaneously detected 
in the waiting mode of the Konus-\textit{Wind} experiment and in the triggered mode of the \BAT experiment.
The sample covers the period from 2005 January to 2018 December. 
By taking advantage of the high sensitivity of \BAT and the wide spectral band of KW we were able to constrain the peak spectral energies, the broadband energy fluences, and the peak fluxes for the joint KW-BAT sample of 167 weak 
($F_\textrm{peak}\lesssim 3 \times 10^{-6}$~erg~cm$^{-2}$~s$^{-1}$), relatively soft GRBs (including four short bursts).
From the spectral analyses of the sample, we provide the spectral fits with CPL and Band model functions.
We calculated the 10~keV--10~MeV energy fluences and the peak energy fluxes on three time scales, including the GRB rest-frame 64~ms scale.
Based on the GRB redshifts, we estimated the rest-frame, isotropic-equivalent energies ($E_\textrm{iso}$) and peak luminosities ($L_\textrm{iso}$).
For 14 GRBs with reasonably constrained jet breaks we provide the collimation-corrected values of the energetics.

This work extends the sample of KW bursts with known redshifts to 338 GRBs, the largest set of cosmological GRBs studied to date over a broad energy band.
With the sample-wide flux limit of $F_\textrm{peak}\sim 1 \times 10^{-7}$~erg~cm$^{-2}$~s$^{-1}$,
it spans redshifts $0.04 \leq z \leq 9.4$, bolometric isotropic luminosities $L_\mathrm{iso}$ from $\sim 2 \times 10^{48}$~erg~s$^{-1}$ to $\sim 5 \times 10^{54}$~erg~s$^{-1}$,
and isotropic energies $E_\mathrm{iso}$ from $\sim 3 \times 10^{49}$~erg to $\sim 6 \times 10^{54}$~erg; and intrinsic peak energies from $\sim$50~keV to $\sim$7~MeV.

Accounting for instrumental bias and using non-parametric techniques, we estimated the GRB luminosity evolution, 
luminosity and isotropic-energy functions, and the evolution of the GRB formation rate for the whole sample of 316 long KW GRBs with known redshifts. 
The derived luminosity evolution and isotropic energy evolution indices $\delta_\textrm{L} \sim 1.2$ and $\delta_\textrm{E} \sim 1.1$ 
are more shallow than those reported in previous studies and are in agreement with T17. 
The shape of the derived LF is best described by a broken PL function with low- and high-luminosity slopes $\sim -0.3$ and $\sim -1$, respectively. 
The EF is better described by an exponentially cut off PL with the PL index $\sim -0.3$ and a cutoff isotropic energy $\sim 10^{54}$~erg. 
The derived GRBFR nearly traces the SFR at $z \gtrsim 1$. At $z<1$, however, it features an excess over the SFR, that cannot be attributed to an analysis artefact.
Finally, we considered the behavior of the rest-frame GRB parameters in the hardness-intensity planes, 
and confirmed the ``Amati'' and ``Yonetoku'' relations for long GRBs. We confirm the result of T17 that the correction for the jet collimation does not improve these correlations for this sample.

Plots of the GRB light curves and spectral fits can be found at the Ioffe Web site\footnote{\url{http://www.ioffe.ru/LEA/zGRBs/part2/}}.
We hope this catalog will encourage further investigations of GRB physical properties and will contribute to other related studies.

We thank the anonymous referee for helpful comments on the manuscript.
We gratefully thank Takanori Sakamoto and Amy Yarleen Lien for helpful discussions on BAT data processing and for their useful comments on the manuscript.
We acknowledge a stimulating discussion with Valentin Pal'shin during an early stage of this work.
We thank Vah\'e Petrosian and Maria Giovanna Dainotti for the enlightening conversations on the LBEP technique.
We acknowledge the use of the public data from the \textit{Swift} data archive\footnote{http://swift.gsfc.nasa.gov} and the Gamma-Ray Burst Online Index (``GRBOX'')\footnote{http://www.astro.caltech.edu/grbox/grbox.php}.

\textit{Facility:} \facility{\textit{Wind}(Konus), \textit{Swift}(BAT)}

\clearpage
\bibliography{BG_GRBs_with_redshifts}{}
\bibliographystyle{aasjournal}

\clearpage

\clearpage

\clearpage

\clearpage

\clearpage

\clearpage
\begin{deluxetable}{lllrccrr}
\tabletypesize{\scriptsize}
\tablewidth{0pt}
\tablecaption{Joint KW/\textit{Swift}-BAT sample\label{generaltab}}
\tablehead{
\colhead{Burst}
& \colhead{\textit{Swift}}
& \colhead{Trigger}
& \colhead{$z$}
& \colhead{$z$\ type\tablenotemark{b}}
& \colhead{$z$} 
& \colhead{$t_0$}
& \colhead{$T_{100}$} 
\\
\colhead{name}
& \colhead{trigger \#}
& \colhead{time\tablenotemark{a}}
& \colhead{}
& \colhead{}
& \colhead{ref} 
& \colhead{(s)}
& \colhead{(s)}
}
\startdata
GRB~050126 & 103780 & 12:00:54.073 & 1.290 & s & (\ref{Gen:Berger2005}) & -0.240 & 40.000\\
GRB~050219A & 106415 & 12:40:01.049 & 0.211 & s & (\ref{Gen:Rossi2014}) & -5.816 & 27.072\\
GRB~050315 & 111063 & 20:59:42.519 & 1.9500 & s & (\ref{Gen:Berger2005a}) & -56.240 & 105.408\\
GRB~050318 & 111529 & 15:44:37.171 & 1.4436 & s & (\ref{Gen:Berger2005a}) & -0.600 & 32.992\\
GRB~050505 & 117504 & 23:22:21.099 & 4.27 & s & (\ref{Gen:Berger2005b}) & -10.648 & 62.656\\
GRB~050724 & 147478 & 12:34:09.362 & 0.258 & s & (\ref{Gen:Prochaska2005}) & -0.056 & 2.096\\
GRB~050730 & 148225 & 19:58:23.199 & 3.9693 & s & (\ref{Gen:Fynbo2009}) & -61.064 & 181.504\\
GRB~050802 & 148646 & 10:08:02.267 & 1.7102 & s & (\ref{Gen:Fynbo2009}) & -3.408 & 22.976\\
GRB~050803 & 148833 & 19:14:00.344 & 0.422 & s & (\ref{Gen:Bloom2005}) & 59.592 & 138.944\\
GRB~050814 & 150314 & 11:38:56.966 & 5.3 & p & (\ref{Gen:Jakobsson2006}) & -1.272 & 108.544\\
GRB~050826 & 152113 & 06:18:10.289 & 0.296 & s & (\ref{Gen:Mirabal2007}) & -0.208 & 39.224\\
GRB~050904 & 153514 & 01:51:44.290 & 6.295 & s & (\ref{Gen:Kawai2006}) & 19.688 & 195.456\\
GRB~051006 & 158593 & 20:30:33.256 & 1.059 & s & (\ref{Gen:Jakobsson2012}) & -3.472 & 35.264\\
GRB~051111 & 163438 & 05:59:41.479 & 1.54948 & s & (\ref{Gen:Penprase2006}) & -5.872 & 87.040\\
GRB~060111A & 176818 & 04:23:06.123 & 2.32 & s & (\ref{Gen:Perley2013a}) & -0.640 & 13.440\\
GRB~060115 & 177408 & 13:08:00.643 & 3.5328 & s & (\ref{Gen:Fynbo2009}) & -50.192 & 158.336\\
GRB~060116 & 177533 & 08:37:27.233 & 6.60 & p & (\ref{Gen:Grazian2006}) & -55.160 & 94.208\\
GRB~060202 & 179968 & 08:40:55.008 & 0.785 & s & (\ref{Gen:Perley2013}) & -32.632 & 224.512\\
GRB~060206 & 180455 & 04:46:53.273 & 4.0559 & s & (\ref{Gen:Fynbo2009}) & -0.584 & 7.040\\
GRB~060210 & 180977 & 04:58:49.809 & 3.9122 & s & (\ref{Gen:Fynbo2009}) & -229.416 & 353.792\\
GRB~060223A & 192059 & 06:04:23.928 & 4.406 & s & (\ref{Gen:Chary2007}) & -2.040 & 12.096\\
GRB~060306 & 200638 & 00:49:10.626 & 1.551 & s & (\ref{Gen:Perley2013}) & -1.312 & 68.416\\
GRB~060418 & 205851 & 03:06:08.204 & 1.4901 & s & (\ref{Gen:Prochaska2006}) & -19.440 & 160.192\\
GRB~060526 & 211957 & 16:28:29.951 & 3.2213 & s & (\ref{Gen:Fynbo2009}) & -1.344 & 10.176\\
GRB~060602A & 213180 & 21:32:12.464 & 0.787 & s & (\ref{Gen:Hjorth2012}) & 1.072 & 73.792\\
GRB~060607A & 213823 & 05:12:13.352 & 3.0749 & s & (\ref{Gen:Fynbo2009}) & -16.104 & 122.368\\
GRB~060707 & 217704 & 21:30:19.497 & 3.4240 & s & (\ref{Gen:Fynbo2009}) & -51.832 & 106.240\\
GRB~060708 & 217805 & 12:15:59.016 & 1.92 & p & (\ref{Gen:Oates2009}) & -1.304 & 7.936\\
GRB~060714 & 219101 & 15:12:00.283 & 2.7108 & s & (\ref{Gen:Fynbo2009}) & -11.696 & 132.416\\
GRB~060729 & 221755 & 19:12:29.244 & 0.5428 & s & (\ref{Gen:Fynbo2009}) & -0.808 & 125.760\\
GRB~060801 & 222154 & 12:16:15.159 & 1.1304\tablenotemark{c} & s & (\ref{Gen:Berger2007}) & -0.512 & 1.960\\
GRB~060904B & 228006 & 02:31:03.858 & 0.7029 & s & (\ref{Gen:Fynbo2009}) & -0.944 & 28.664\\
GRB~060906 & 228316 & 08:32:46.573 & 3.6856 & s & (\ref{Gen:Fynbo2009}) & -40.000 & 49.448\\
GRB~060908 & 228581 & 08:57:22.345 & 1.8836 & s & (\ref{Gen:Fynbo2009}) & -10.208 & 22.656\\
GRB~060923A & 230662 & 05:12:15.370 & 2.6 & p & (\ref{Gen:Perley2013}) & -42.264 & 56.384\\
GRB~060927 & 231362 & 14:07:35.297 & 5.4636 & s & (\ref{Gen:Fynbo2009}) & -1.112 & 25.408\\
GRB~061110B & 238174 & 21:58:45.541 & 3.4344 & s & (\ref{Gen:Fynbo2009}) & -16.256 & 147.578\\
GRB~070110 & 255445 & 07:22:41.574 & 2.3521 & s & (\ref{Gen:Fynbo2009}) & -2.784 & 93.632\\
GRB~070208 & 259714 & 09:10:34.281 & 1.165 & s & (\ref{Gen:Cucchiara2007}) & -4.032 & 55.176\\
GRB~070318 & 271019 & 07:28:56.089 & 0.8397 & s & (\ref{Gen:Fynbo2009}) & -0.880 & 59.809\\
GRB~070411 & 275087 & 20:12:33.316 & 2.9538 & s & (\ref{Gen:Fynbo2009}) & -16.416 & 107.904\\
GRB~070419B & 276212 & 10:44:05.972 & 1.9588 & s & (\ref{Gen:Kruehler2012b}) & -12.608 & 275.136\\
GRB~070529 & 280706 & 12:48:28.349 & 2.4996 & s & (\ref{Gen:Berger2007a}) & 0.664 & 120.192\\
GRB~070611 & 282003 & 01:57:13.890 & 2.0394 & s & (\ref{Gen:Fynbo2009}) & -3.232 & 10.560\\
GRB~070612A & 282066 & 02:38:45.984 & 0.617 & s & (\ref{Gen:Cenko2007}) & -5.976 & 281.664\\
GRB~070721B & 285654 & 10:33:46.314 & 3.6298 & s & (\ref{Gen:Fynbo2009}) & -6.712 & 371.136\\
GRB~070810A & 287364 & 02:11:52.415 & 2.17 & s & (\ref{Gen:Thoene2007}) & -2.120 & 6.400\\
GRB~071021 & 294974 & 09:41:33.694 & 2.4520 & s & (\ref{Gen:Kruehler2012b}) & -4.504 & 264.768\\
GRB~071025 & 295301 & 04:08:53.685 & 5.2\tablenotemark{d} & s+p & (\ref{Gen:Fynbo2009}) & 40.104 & 152.704\\
GRB~080129 & 301981 & 06:06:45.464 & 4.349 & s & (\ref{Gen:Greiner2009}) & -1.704 & 38.976\\
GRB~080207 & 302728 & 21:30:21.442 & 2.0858 & s & (\ref{Gen:Kruehler2012b}) & -1.112 & 343.112\\
GRB~080210 & 302888 & 07:50:05.436 & 2.6419 & s & (\ref{Gen:Fynbo2009}) & -14.560 & 62.400\\
GRB~080310 & 305288 & 08:37:58.647 & 2.4274 & s & (\ref{Gen:Fynbo2009}) & -9.712 & 21.765\\
GRB~080325 & 307604 & 04:09:17.332 & 1.78 & s & (\ref{Gen:Perley2013}) & -33.464 & 213.504\\
GRB~080430 & 310613 & 19:53:02.076 & 0.767 & s & (\ref{Gen:Cucchiara2008}) & -0.408 & 20.352\\
GRB~080515 & 311658 & 06:01:13.817 & 2.47 & s & (\ref{Gen:Perley2013a}) & -6.400 & 24.792\\
GRB~080516 & 311762 & 00:17:07.031 & 3.2 & p & (\ref{Gen:Filgas2008}) & 0.048 & 7.040\\
GRB~080604 & 313116 & 07:27:01.160 & 1.4171 & s & (\ref{Gen:Fynbo2009}) & -28.632 & 88.768\\
GRB~080707 & 316204 & 08:27:53.686 & 1.2322 & s & (\ref{Gen:Fynbo2009}) & -2.440 & 31.360\\
GRB~080804 & 319016 & 23:20:14.670 & 2.2045 & s & (\ref{Gen:Fynbo2009}) & -0.496 & 38.784\\
GRB~080805 & 319036 & 07:41:34.733 & 1.5042 & s & (\ref{Gen:Fynbo2009}) & -3.888 & 96.832\\
GRB~080810 & 319584 & 13:10:12.288 & 3.3604 & s & (\ref{Gen:Fynbo2009}) & -14.016 & 120.256\\
GRB~080905B & 323898 & 16:55:45.405 & 2.3739 & s & (\ref{Gen:Fynbo2009}) & -7.480 & 111.872\\
GRB~080906 & 323984 & 13:33:16.347 & 2.13 & p & (\ref{Gen:Kruehler2011a}) & -117.888 & 235.008\\
GRB~080913 & 324561 & 06:46:54.122 & 6.7 & s & (\ref{Gen:Fynbo2008}) & -3.592 & 8.832\\
GRB~080928 & 326115 & 15:01:32.867 & 1.6919 & s & (\ref{Gen:Fynbo2009}) & 56.808 & 268.352\\
GRB~081008 & 331093 & 19:58:09.385 & 1.9685 & s & (\ref{Gen:DAvanzo2008}) & -65.160 & 207.104\\
GRB~081028A & 332851 & 00:25:00.791 & 3.038 & s & (\ref{Gen:Berger2008a}) & 44.656 & 300.608\\
GRB~081029 & 332931 & 01:43:56.788 & 3.8479 & s & (\ref{Gen:DElia2008}) & -38.864 & 326.528\\
GRB~081109A & 334112 & 07:02:06.615 & 0.9787 & s & (\ref{Gen:Kruehler2011}) & -18.096 & 95.488\\
GRB~090113 & 339852 & 18:40:39.171 & 1.7493 & s & (\ref{Gen:Kruehler2012b}) & -0.160 & 10.112\\
GRB~090418A & 349510 & 11:07:40.227 & 1.608 & s & (\ref{Gen:Chornock2009a}) & -8.216 & 69.184\\
GRB~090426 & 350479 & 12:48:47.213 & 2.609 & s & (\ref{Gen:Levesque2009}) & -0.032 & 0.832\\
GRB~090429B & 350854 & 05:30:03.366 & 9.38 & p & (\ref{Gen:Cucchiara2011a}) & -3.064 & 6.000\\
GRB~090516A & 352190 & 08:27:50.774 & 4.109 & s & (\ref{Gen:deUgartePostigo2009}) & -8.544 & 234.560\\
GRB~090519 & 352648 & 21:08:56.429 & 3.85 & s & (\ref{Gen:Levan2009}) & -12.432 & 86.976\\
GRB~090530 & 353567 & 03:18:18.384 & 1.266 & s & (\ref{Gen:Goldoni2013}) & -0.224 & 41.600\\
GRB~090726 & 358422 & 22:42:27.759 & 2.71 & s & (\ref{Gen:Fatkhullin2009}) & -22.879 & 61.824\\
GRB~090814A & 359951 & 00:52:19.004 & 0.696\tablenotemark{e} & s & (\ref{Gen:Jakobsson2009}) & -15.200 & 61.248\\
GRB~090926B & 370791 & 21:55:48.416 & 1.24 & s & (\ref{Gen:Fynbo2009a}) & -21.352 & 154.496\\
GRB~091018 & 373172 & 20:48:19.585 & 0.971 & s & (\ref{Gen:Chen2009}) & -0.376 & 5.184\\
GRB~091029 & 374210 & 03:53:22.596 & 2.752 & s & (\ref{Gen:Chornock2009}) & -10.176 & 50.432\\
GRB~091109A & 375246 & 04:57:43.364 & 3.076 & s & (\ref{Gen:Rau2010}) & -0.824 & 24.832\\
GRB~100316A & 416076 & 02:23:00.430 & 3.155 & s & (\ref{Gen:Sanchez-Ramirez2013}) & -1.032 & 7.040\\
GRB~100316B & 416103 & 08:01:36.973 & 1.180 & s & (\ref{Gen:Vergani2010}) & -0.360 & 4.672\\
GRB~100615A & 424733 & 01:59:03.995 & 1.398 & s & (\ref{Gen:Kruehler2013}) & -0.080 & 47.488\\
GRB~100728B & 430172 & 10:31:55.829 & 2.106 & s & (\ref{Gen:Flores2010}) & -1.712 & 14.912\\
GRB~100901A & 433065 & 13:34:10.376 & 1.408 & s & (\ref{Gen:Chornock2010}) & -2.776 & 444.800\\
GRB~101219B & 440635 & 16:27:53.520 & 0.5519 & s & (\ref{Gen:deUgartePostigo2011}) & 15.720 & 26.496\\
GRB~110106B & 441676 & 21:26:17.014 & 0.618 & s & (\ref{Gen:Chornock2011a}) & -3.008 & 24.448\\
GRB~110128A & 443861 & 01:44:33.009 & 2.339 & s & (\ref{Gen:Sparre2011}) & -2.560 & 24.320\\
GRB~110205A & 444643 & 02:02:41.367 & 2.22 & s & (\ref{Gen:Cenko2011}) & 21.416 & 356.544\\
GRB~110726A & 458059 & 01:30:40.550 & 1.036\tablenotemark{f} & s & (\ref{Gen:Cucchiara2011}) & -0.064 & 3.776\\
GRB~110801A & 458521 & 19:49:42.991 & 1.858 & s & (\ref{Gen:CabreraLavers2011}) & -24.232 & 412.160\\
GRB~110818A & 500914 & 20:37:49.206 & 3.36 & s & (\ref{Gen:DAvanzo2011}) & -13.048 & 77.440\\
GRB~111107A & 507185 & 00:50:24.005 & 2.893 & s & (\ref{Gen:Chornock2011}) & -0.064 & 37.440\\
GRB~111123A & 508319 & 18:13:21.100 & 3.1516 & s & (\ref{Gen:Xu2013a}) & -8.216 & 286.336\\
GRB~111225A & 510341 & 03:50:37.792 & 0.297 & s & (\ref{Gen:Thoene2014}) & -8.720 & 82.496\\
GRB~120118B & 512003 & 17:00:21.195 & 2.943 & s & (\ref{Gen:Malesani2013}) & -0.824 & 26.496\\
GRB~120326A & 518626 & 01:20:29.280 & 1.798 & s & (\ref{Gen:Tello2012}) & -70.992 & 88.576\\
GRB~120327A & 518731 & 02:55:16.636 & 2.813 & s & (\ref{Gen:Kruehler2012a}) & -15.992 & 83.136\\
GRB~120404A & 519380 & 12:51:02.399 & 2.876 & s & (\ref{Gen:Cucchiara2012}) & -5.096 & 42.624\\
GRB~120521C & 522656 & 23:22:07.703 & 6 & p & (\ref{Gen:Laskar2014}) & -1.472 & 38.400\\
GRB~120712A & 526351 & 13:42:27.382 & 4.1745 & s & (\ref{Gen:Xu2012}) & -4.600 & 22.272\\
GRB~120722A & 528195 & 12:53:26.546 & 0.9586 & s & (\ref{Gen:DElia2012}) & -2.968 & 38.208\\
GRB~120802A & 529486 & 08:00:51.628 & 3.796 & s & (\ref{Gen:Tanvir2012a}) & -1.040 & 17.984\\
GRB~120811C & 530689 & 15:34:52.170 & 2.671 & s & (\ref{Gen:Thoene2012}) & -9.656 & 34.688\\
GRB~120815A & 531003 & 02:13:58.782 & 2.358 & s & (\ref{Gen:Malesani2012}) & -0.216 & 7.872\\
GRB~120907A & 532871 & 00:24:23.082 & 0.970 & s & (\ref{Gen:Sanchez-Ramirez2012a}) & -0.024 & 7.104\\
GRB~120909A & 533060 & 01:42:03.235 & 3.93 & s & (\ref{Gen:Hartoog2012}) & -40.512 & 125.760\\
GRB~120922A & 534394 & 22:30:28.657 & 3.1 & p & (\ref{Gen:Knust2012}) & -22.568 & 171.776\\
GRB~121024A & 536580 & 02:56:12.478 & 2.298 & s & (\ref{Gen:Tanvir2012}) & -8.296 & 12.096\\
GRB~121027A & 536831 & 07:32:29.747 & 1.773 & s & (\ref{Gen:Kruehler2012}) & -9.344 & 72.640\\
GRB~121201A & 540178 & 12:25:42.065 & 3.385 & s & (\ref{Gen:Sanchez-Ramirez2012}) & -24.088 & 7.008\\
GRB~121211A & 541200 & 13:47:02.795 & 1.023 & s & (\ref{Gen:Perley2012}) & -2.776 & 199.936\\
GRB~130131B & 547420 & 19:10:08.907 & 2.539 & s & (\ref{Gen:Fynbo2013}) & -0.288 & 4.736\\
GRB~130420A & 553977 & 07:28:29.538 & 1.297 & s & (\ref{Gen:deUgartePostigo2013}) & -24.728 & 207.360\\
GRB~130427B & 554635 & 13:20:41.799 & 2.78\tablenotemark{g} & s & (\ref{Gen:Flores2013}) & -0.632 & 30.144\\
GRB~130511A & 555600 & 11:30:47.442 & 1.3033 & s & (\ref{Gen:Cucchiara2013a}) & -0.048 & 3.328\\
GRB~130514A & 555821 & 07:13:41.063 & 3.6 & p & (\ref{Gen:Schmidl2013}) & -7.680 & 281.344\\
GRB~130604A & 557354 & 06:54:26.998 & 1.06 & s & (\ref{Gen:Cenko2013}) & -0.624 & 34.816\\
GRB~130606A & 557589 & 21:04:39.020 & 5.91 & s & (\ref{Gen:Castro-Tirado2013}) & -1.104 & 165.970\\
GRB~130610A & 557845 & 03:12:13.344 & 2.092 & s & (\ref{Gen:Smette2013}) & -6.688 & 67.688\\
GRB~130612A & 557976 & 03:22:22.168 & 2.006 & s & (\ref{Gen:Tanvir2013}) & -0.504 & 3.264\\
GRB~131004A & 573190 & 21:41:03.688 & 0.717 & s & (\ref{Gen:Chornock2013}) & -0.248 & 1.344\\
GRB~131103A & 576562 & 22:07:25.794 & 0.5955 & s & (\ref{Gen:Xu2013}) & -8.784 & 15.760\\
GRB~131227A & 582184 & 04:44:51.212 & 5.3 & s & (\ref{Gen:Cucchiara2013}) & -1.064 & 20.288\\
GRB~140114A & 583861 & 11:57:40.482 & 3.0 & s & (\ref{Gen:Kruehler2015}) & 24.296 & 141.056\\
GRB~140304A & 590206 & 13:22:31.098 & 5.283 & s & (\ref{Gen:Jeong2014}) & -3.912 & 19.456\\
GRB~140311A & 591390 & 21:05:16.252 & 4.954 & s & (\ref{Gen:Chornock2014a}) & -4.440 & 85.184\\
GRB~140423A & 596901 & 08:31:53.262 & 3.26 & s & (\ref{Gen:Tanvir2014a}) & -67.472 & 182.080\\
GRB~140430A & 597722 & 20:33:36.527 & 1.60 & s & (\ref{Gen:Kruehler2014a}) & -0.344 & 29.864\\
GRB~140509A & 598497 & 02:22:13.613 & 2.4 & p & (\ref{Gen:Marshall2014}) & -3.312 & 29.312\\
GRB~140518A & 599287 & 09:17:46.631 & 4.707 & s & (\ref{Gen:Chornock2014}) & -4.632 & 64.704\\
GRB~140614A & 601646 & 01:04:59.864 & 4.233 & s & (\ref{Gen:Kruehler2014}) & -96.600 & 184.768\\
GRB~140629A & 602884 & 14:17:30.327 & 2.275 & s & (\ref{Gen:Moskvitin2014}) & -6.184 & 26.304\\
GRB~140703A & 603243 & 00:37:17.064 & 3.14 & s & (\ref{Gen:Castro-Tirado2014a}) & -14.128 & 86.208\\
GRB~140710A & 603954 & 10:16:40.041 & 0.558 & s & (\ref{Gen:Tanvir2014}) & -0.176 & 1.728\\
GRB~140907A & 611933 & 16:07:08.847 & 1.21 & s & (\ref{Gen:Castro-Tirado2014}) & -16.176 & 55.552\\
GRB~141004A & 614390 & 23:20:54.395 & 0.573 & s & (\ref{Gen:deUgartePostigo2014}) & -0.864 & 6.208\\
GRB~141109A & 618024 & 05:49:55.236 & 2.993 & s & (\ref{Gen:Xu2014}) & -0.664 & 179.584\\
GRB~141225A & 622476 & 23:01:07.035 & 0.915 & s & (\ref{Gen:Gorosabel2014}) & 4.904 & 28.352\\
GRB~150301B & 633180 & 19:38:04.028 & 1.5169 & s & (\ref{Gen:deUgartePostigo2015a}) & -0.328 & 24.384\\
GRB~150413A & 637899 & 13:54:58.559 & 3.139 & s & (\ref{Gen:deUgartePostigo2015}) & -90.648 & 300.352\\
GRB~150818A & 652603 & 11:36:32.926 & 0.282 & s & (\ref{Gen:Sanchez-Ramirez2015}) & -26.112 & 184.256\\
GRB~150910A & 655097 & 09:04:48.886 & 1.359 & s & (\ref{Gen:Zheng2015}) & 32.616 & 90.752\\
GRB~151111A & 663074 & 08:33:23.417 & 3.5 & p & (\ref{Gen:Bolmer2015a}) & -5.128 & 90.752\\
GRB~151112A & 663179 & 13:44:48.084 & 4.1 & p & (\ref{Gen:Bolmer2015}) & -0.064 & 14.784\\
GRB~151215A & 667392 & 03:01:28.957 & 2.59 & s & (\ref{Gen:Xu2015}) & -0.216 & 3.264\\
GRB~160121A & 671231 & 13:50:37.711 & 1.960 & s & (\ref{Gen:Sanchez-Ramirez2016}) & -0.016 & 12.864\\
GRB~160227A & 676423 & 19:32:08.096 & 2.38 & s & (\ref{Gen:Xu2016}) & -20.056 & 287.168\\
GRB~160327A & 680655 & 09:16:07.705 & 4.99 & s & (\ref{Gen:deUgartePostigo2016}) & -6.400 & 42.176\\
GRB~161017A & 718023 & 17:51:51.170 & 2.013 & s & (\ref{Gen:Castro-Tirado2016}) & -4.632 & 240.832\\
GRB~161108A & 721234 & 03:32:33.177 & 1.159 & s & (\ref{Gen:deUgartePostigo2016a}) & -4.296 & 130.048\\
GRB~161219B & 727541 & 18:48:39.308 & 0.1475 & s & (\ref{Gen:Tanvir2016}) & -0.536 & 7.040\\
GRB~170113A & 732526 & 10:04:05.482 & 1.968 & s & (\ref{Gen:Xu2017}) & -0.608 & 23.672\\
GRB~170202A & 736407 & 18:28:02.373 & 3.645 & s & (\ref{Gen:deUgartePostigo2017}) & 0.328 & 50.744\\
GRB~170531B & 755354 & 22:02:09.206 & 2.366 & s & (\ref{Gen:deUgartePostigo2017a}) & -0.056 & 171.904\\
GRB~170604A & 755867 & 19:08:50.402 & 1.329 & s & (\ref{Gen:Izzo2017a}) & -11.672 & 30.592\\
GRB~171020A & 780845 & 23:07:10.752 & 1.87 & s & (\ref{Gen:Malesani2017}) & -0.968 & 28.352\\
GRB~171205A & 794972 & 07:20:43.893 & 0.0368 & s & (\ref{Gen:Izzo2017}) & -44.120 & 216.896\\
GRB~171222A & 799669 & 16:24:59.804 & 2.409 & s & (\ref{Gen:deUgartePostigo2017b}) & 4.968 & 205.120\\
GRB~180115A & 805318 & 04:16:03.681 & 2.487 & s & (\ref{Gen:deUgartePostigo2018}) & 16.680 & 56.128\\
GRB~180205A & 808625 & 04:25:29.332 & 1.409 & s & (\ref{Gen:Tanvir2018}) & -6.992 & 13.120\\
GRB~180329B & 819490 & 14:08:23.971 & 1.998 & s & (\ref{Gen:Izzo2018}) & -8.888 & 266.496\\
GRB~180404A & 821881 & 00:45:35.687 & 1.000 & s & (\ref{Gen:Selsing2018}) & -22.712 & 38.400\\
GRB~180624A & 844192 & 13:49:40.379 & 2.855 & s & (\ref{Gen:Rossi2018}) & -61.128 & 262.080\\
GRB~181110A & 871316 & 08:43:31.320 & 1.505 & s & (\ref{Gen:Perley2018}) & -104.432 & 217.536\\
\enddata
\tablenotetext{a}{\textit{Swift}/BAT trigger time.}
\tablenotetext{b}{Redshift types are: s = spectroscopic, p = photometric.}
\tablenotetext{c}{There are two sources located within the revised XRT error circle. Here we use the redshift of one of them as a proxy for the GRB redshift. The other source, which is a factor of 2 fainter, is likely to reside at an even higher redshift.}
\tablenotetext{d}{No emission was detected at wavelengths shorter than $ 7500 \AA $. This flux ``decrement'' may be associated with the IGM at $ z \approx 5.2 $ but could also be associated with a significant reddening of the afterglow.}
\tablenotetext{e}{A weak absorption system identified at z=0.696 could also be produced by an intervening system and thus the strict redshift range for this GRB would be $0.696 < z < 2.2$.}
\tablenotetext{f}{The object is well detected and presents a featureless continuum except for a weak double line at $ z = 1.036 $. No other clear features are detected. Therefore, $1.036 < z < 2.7$ is suggested as redshift range for this GRB.}\tablenotetext{g}{Given the low S/N of the spectrum, this redshift measurement should be considered tentative.}\tablerefs{\cititem{Berger2005}; \cititem{Rossi2014}; \cititem{Berger2005a}; \cititem{Berger2005b}; \cititem{Prochaska2005}; \cititem{Fynbo2009}; \cititem{Bloom2005}; \cititem{Jakobsson2006}; \cititem{Mirabal2007}; \cititem{Kawai2006}; \cititem{Jakobsson2012}; \cititem{Penprase2006}; \cititem{Perley2013a}; \cititem{Grazian2006}; \cititem{Perley2013}; \cititem{Chary2007}; \cititem{Prochaska2006}; \cititem{Hjorth2012}; \cititem{Oates2009}; \cititem{Berger2007}; \cititem{Cucchiara2007}; \cititem{Kruehler2012b}; \cititem{Berger2007a}; \cititem{Cenko2007}; \cititem{Thoene2007}; \cititem{Greiner2009}; \cititem{Cucchiara2008}; \cititem{Filgas2008}; \cititem{Kruehler2011a}; \cititem{Fynbo2008}; \cititem{DAvanzo2008}; \cititem{Berger2008a}; \cititem{DElia2008}; \cititem{Kruehler2011}; \cititem{Chornock2009a}; \cititem{Levesque2009}; \cititem{Cucchiara2011a}; \cititem{deUgartePostigo2009}; \cititem{Levan2009}; \cititem{Goldoni2013}; \cititem{Fatkhullin2009}; \cititem{Jakobsson2009}; \cititem{Fynbo2009a}; \cititem{Chen2009}; \cititem{Chornock2009}; \cititem{Rau2010}; \cititem{Sanchez-Ramirez2013}; \cititem{Vergani2010}; \cititem{Kruehler2013}; \cititem{Flores2010}; \cititem{Chornock2010}; \cititem{deUgartePostigo2011}; \cititem{Chornock2011a}; \cititem{Sparre2011}; \cititem{Cenko2011}; \cititem{Cucchiara2011}; \cititem{CabreraLavers2011}; \cititem{DAvanzo2011}; \cititem{Chornock2011}; \cititem{Xu2013a}; \cititem{Thoene2014}; \cititem{Malesani2013}; \cititem{Tello2012}; \cititem{Kruehler2012a}; \cititem{Cucchiara2012}; \cititem{Laskar2014}; \cititem{Xu2012}; \cititem{DElia2012}; \cititem{Tanvir2012a}; \cititem{Thoene2012}; \cititem{Malesani2012}; \cititem{Sanchez-Ramirez2012a}; \cititem{Hartoog2012}; \cititem{Knust2012}; \cititem{Tanvir2012}; \cititem{Kruehler2012}; \cititem{Sanchez-Ramirez2012}; \cititem{Perley2012}; \cititem{Fynbo2013}; \cititem{deUgartePostigo2013}; \cititem{Flores2013}; \cititem{Cucchiara2013a}; \cititem{Schmidl2013}; \cititem{Cenko2013}; \cititem{Castro-Tirado2013}; \cititem{Smette2013}; \cititem{Tanvir2013}; \cititem{Chornock2013}; \cititem{Xu2013}; \cititem{Cucchiara2013}; \cititem{Kruehler2015}; \cititem{Jeong2014}; \cititem{Chornock2014a}; \cititem{Tanvir2014a}; \cititem{Kruehler2014a}; \cititem{Marshall2014}; \cititem{Chornock2014}; \cititem{Kruehler2014}; \cititem{Moskvitin2014}; \cititem{Castro-Tirado2014a}; \cititem{Tanvir2014}; \cititem{Castro-Tirado2014}; \cititem{deUgartePostigo2014}; \cititem{Xu2014}; \cititem{Gorosabel2014}; \cititem{deUgartePostigo2015a}; \cititem{deUgartePostigo2015}; \cititem{Sanchez-Ramirez2015}; \cititem{Zheng2015}; \cititem{Bolmer2015a}; \cititem{Bolmer2015}; \cititem{Xu2015}; \cititem{Sanchez-Ramirez2016}; \cititem{Xu2016}; \cititem{deUgartePostigo2016}; \cititem{Castro-Tirado2016}; \cititem{deUgartePostigo2016a}; \cititem{Tanvir2016}; \cititem{Xu2017}; \cititem{deUgartePostigo2017}; \cititem{deUgartePostigo2017a}; \cititem{Izzo2017a}; \cititem{Malesani2017}; \cititem{Izzo2017}; \cititem{deUgartePostigo2017b}; \cititem{deUgartePostigo2018}; \cititem{Tanvir2018}; \cititem{Izzo2018}; \cititem{Selsing2018}; \cititem{Rossi2018}; \cititem{Perley2018}; }

\end{deluxetable}

\begin{deluxetable}{lccrcrcrrc}
\rotate
\tabletypesize{\scriptsize}
\tablewidth{0pt}
\tablecaption{Spectral Parameters \label{spectraltab}}
\tablehead{
\colhead{Burst}
& \colhead{Spec.}
& \colhead{$t_\textrm{start}$}
& \colhead{$\Delta T$}
& \colhead{Model}
& \colhead{$\alpha$}
& \colhead{$\beta$}
& \colhead{$E_\textrm{p}$}
& \colhead{$F$}
& \colhead{$\chi ^{2}/\textrm{d.o.f.}$} 
\\
\colhead{name}
& \colhead{type}
& \colhead{(s)}
& \colhead{(s)}
& \colhead{}
& \colhead{}
& \colhead{}
& \colhead{(keV)}
& \colhead{($10^{-7}$~erg~cm$^{-2}$~s$^{-1}$)}
& \colhead{(Prob.)} 
}
\startdata
GRB~050126 & i & $ -3.168 $ & $ 38.272 $ & CPL & $ -0.90_{-0.23}^{+0.28} $ & $ \dots $ & $ 158_{-41}^{+82} $ & $ 0.63_{-0.12}^{+0.19} $ & $ 72.1/59 \: (0.12) $\\
 & p & $ -0.224 $ & $ 5.888 $ & CPL &$ -0.78_{-0.27}^{+0.37} $ & $ \dots $ & $ 239_{-92}^{+218} $ & $1.29_{-0.41}^{+0.68}$ & $55.5/59 \: (0.61)$\\
GRB~050219A & i & $ -5.223 $ & $ 26.496 $ & BAND & $ 0.02_{-0.17}^{+0.19} $ & $ -3.35_{-6.65}^{+0.55} $ & $ 93_{-5}^{+6} $ & $ 2.33_{-0.25}^{+0.27} $ & $ 44.5/58 \: (0.9) $\\
 & & &  & CPL &$ -0.04_{-0.14}^{+0.15} $ & $ \dots $ & $ 96_{-4}^{+4} $ & $ 2.12_{-0.07}^{+0.08} $ & $ 45.0/59 \: (0.91) $\\
 & p & $ 6.553 $ & $ 5.888 $ & CPL &$ 0.05_{-0.17}^{+0.18} $ & $ \dots $ & $ 110_{-6}^{+7} $ & $4.02_{-0.19}^{+0.21}$ & $41.6/59 \: (0.96)$\\
GRB~050315 & i & $ -57.582 $ & $ 85.376 $ & CPL & $ -1.41_{-0.18}^{+0.19} $ & $ \dots $ & $ 47_{-5}^{+4} $ & $ 0.59_{-0.03}^{+0.04} $ & $ 62.2/59 \: (0.36) $\\
 & p & $ -1.646 $ & $ 8.832 $ & BAND &$ -1.48_{-0.17}^{+0.22} $ & $ -2.39_{-7.61}^{+0.41} $ & $ 113_{-32}^{+81} $ & $2.33_{-0.45}^{+0.53}$ & $66.3/58 \: (0.21)$\\
 & & &  & CPL &$ -1.54_{-0.14}^{+0.18} $ & $ \dots $ & $ 132_{-37}^{+108} $ & $ 2.12_{-0.33}^{+0.52} $ & $ 66.5/59 \: (0.24) $\\
GRB~050318 & i & $ -0.243 $ & $ 35.328 $ & BAND & $ 0.11_{-1.03}^{+1.47} $ & $ -2.24_{-0.18}^{+0.10} $ & $ 34_{-5}^{+8} $ & $ 0.71_{-0.07}^{+0.07} $ & $ 43.1/58 \: (0.93) $\\
 & & &  & CPL &$ -1.21_{-0.21}^{+0.23} $ & $ \dots $ & $ 48_{-4}^{+4} $ & $ 0.47_{-0.03}^{+0.04} $ & $ 45.4/59 \: (0.9) $\\
 & p & $ 23.309 $ & $ 8.832 $ & BAND &$ -1.17_{-0.19}^{+0.24} $ & $ -2.60_{-0.99}^{+0.29} $ & $ 55_{-5}^{+5} $ & $1.75_{-0.28}^{+0.27}$ & $55.1/58 \: (0.58)$\\
 & & &  & CPL &$ -1.23_{-0.16}^{+0.17} $ & $ \dots $ & $ 57_{-4}^{+5} $ & $ 1.45_{-0.08}^{+0.10} $ & $ 56.5/59 \: (0.57) $\\
\enddata

\tablecomments{This table is available in its entirety in machine-readable form.}
\end{deluxetable}

\setcounter{cit}{0}
\renewcommand{\cititem}[1]{\refstepcounter{cit}(\arabic{cit})~\label{En:#1}\citealt{#1}}

\begin{deluxetable}{lcccccccc}
\tabletypesize{\scriptsize}
\tablewidth{0pt}
\tablecaption{Statistics\label{stattab}}
\tablehead {
\colhead{Parameter}
& \colhead{Min\tablenotemark{a}}
& \colhead{Max\tablenotemark{a}}
& \colhead{Mean\tablenotemark{a}}
& \colhead{Median\tablenotemark{a}}
& \colhead{Min\tablenotemark{b}}
& \colhead{Max\tablenotemark{b}}
& \colhead{Mean\tablenotemark{b}}
& \colhead{Median\tablenotemark{b}}
\\
\colhead{}
& \colhead{}
& \colhead{}
& \colhead{}
& \colhead{}
& \colhead{}
& \colhead{}
& \colhead{}
& \colhead{}
}
\startdata
Redshift & 0.04 & 9.4 & 2.4 & 2.3 & 0.04 & 9.4 & 2.0 & 1.7\\
$E_\textrm{p,i}$~(keV) & 27 & 578 & 144 & 111 & 27 & 3670 & 238 & 166\\
$E_\textrm{p,p}$~(keV) & 27 & 578 & 171 & 134 & 27 & 3520 & 282 & 199\\
$E_\textrm{p,i,z}$~(keV) & 53 & 2870 & 509 & 313 & 53 & 6970 & 650 & 504\\
$E_\textrm{p,p,z}$~(keV) & 53 & 2870 & 596 & 425 & 53 & 6680 & 768 & 570\\
$S$ ($10^{-6}$~erg~cm$^{-2}$)  & 0.1 & 40.5 & 6.0 & 3.9 & 0.1 & 3028 & 62.3 & 8.0\\
$F_\textrm{peak,64,r}$ ($10^{-6}$~erg~cm$^{-2}$~s$^{-1}$)  & 0.1 & 2.9 & 0.5 & 0.4 & 0.1 & 898 & 12.9 & 1.2\\
$E_\textrm{iso}$ ($10^{51}$~erg)  & 0.03 & 1477 & 133 & 60 & 0.03 & 5925 & 363 & 86\\
$L_\textrm{iso}$ ($10^{51}$~erg~s$^{-1}$)  & 0 & 514 & 41 & 18 & 0.002 & 4632 & 146 & 32\\
\hline
$\theta_\textrm{HM}$ (deg)  & 1.3 & 10.2 & 4.0 & 3.8 & 1.3 & 25.4 & 5.8 & 4.5\\
Coll. factor HM ($\times 10^{-3}$)  & 0.2 & 15.8 & 3.3 & 2.2 & 0.2 & 96.4 & 7.6 & 3.0\\
$E_{\gamma \textrm{,HM}}$ ($10^{50}$~erg)  & 0.06 & 23.2 & 5.2 & 1.5 & 0.06 & 159.9 & 15.9 & 4.7\\
$L_{\gamma \textrm{,HM}}$ ($10^{50}$~erg~s$^{-1}$)  & 0.04 & 2.9 & 1.0 & 0.5 & 0.04 & 56.0 & 4.1 & 1.8\\
\enddata

\tablenotetext{a} {For the sample from this paper.}
\tablenotetext{b} {For the samples from this paper and T17.}
\end{deluxetable}

\begin{deluxetable}{llrrrrrrrrr}
\rotate
\tabletypesize{\scriptsize} 
\tablewidth{0pt}
\tablecaption{Burst Energetics\label{energytab}}
\tablehead {
\colhead{Burst name}
& \colhead{$z$}
& \colhead{$S$\tablenotemark{a}}
& \colhead{$T_{\rm peak,1000}$\tablenotemark{b}}
& \colhead{$F_{\rm peak,1000}$\tablenotemark{c}}
& \colhead{$T_{\rm peak,64}$\tablenotemark{b}}
& \colhead{$F_{\rm peak,64}$\tablenotemark{c}}
& \colhead{$T_{\rm peak,64,r}$\tablenotemark{b}}
& \colhead{$F_{\rm peak,64,r}$\tablenotemark{c}}
& \colhead{$E_{\rm iso}$\tablenotemark{d}}
& \colhead{$L_{\rm iso}$\tablenotemark{e}} 
}
\startdata
GRB~050126 & $1.290$ & $24.70_{-5.14}^{+7.87}$ & $4.560$ & $1.46_{-0.54}^{+0.82}$ & $3.600$ & $4.13_{-1.80}^{+2.50}$ & $4.525$ & $2.37_{-1.07}^{+1.47}$ & $12.15_{-2.53}^{+3.87}$ & $2.57_{-1.16}^{+1.59}$ \\
GRB~050219A & $0.211$ & $62.70_{-7.04}^{+7.56}$ & $9.552$ & $3.91_{-0.32}^{+0.33}$ & $9.544$ & $7.17_{-1.28}^{+1.29}$ & $9.688$ & $6.65_{-1.09}^{+1.10}$ & $0.76_{-0.09}^{+0.09}$ & $0.10_{-0.02}^{+0.02}$ \\
GRB~050315 & $1.9500$ & $56.90_{-3.93}^{+4.41}$ & $4.048$ & $2.48_{-0.54}^{+0.63}$ & $4.496$ & $5.95_{-1.59}^{+1.76}$ & $4.479$ & $3.39_{-0.91}^{+1.00}$ & $86.89_{-5.99}^{+6.73}$ & $13.99_{-3.75}^{+4.14}$ \\
GRB~050318 & $1.4436$ & $24.90_{-3.43}^{+3.63}$ & $28.392$ & $2.65_{-0.45}^{+0.43}$ & $28.904$ & $5.04_{-1.01}^{+0.99}$ & $28.796$ & $4.78_{-0.86}^{+0.84}$ & $18.39_{-2.53}^{+2.68}$ & $8.90_{-1.61}^{+1.57}$ \\
GRB~050505 & $4.27$ & $47.40_{-5.43}^{+6.73}$ & $0.392$ & $2.97_{-0.64}^{+0.84}$ & $3.496$ & $5.39_{-1.65}^{+1.93}$ & $0.908$ & $3.39_{-0.81}^{+1.03}$ & $192.30_{-22.07}^{+27.31}$ & $72.26_{-17.35}^{+22.02}$ \\
GRB~050724 & $0.258$ & $9.66_{-2.72}^{+2.62}$ & $0.000$ & $6.83_{-1.92}^{+1.85}$ & $0.008$ & $31.90_{-9.09}^{+8.77}$ & $0.025$ & $28.70_{-8.20}^{+7.91}$ & $0.24_{-0.07}^{+0.06}$ & $0.88_{-0.25}^{+0.24}$ \\
GRB~050730 & $3.9693$ & $187.00_{-69.40}^{+88.70}$ & $5.152$ & $3.95_{-1.58}^{+1.97}$ & $9.592$ & $11.90_{-5.03}^{+6.14}$ & $9.553$ & $6.28_{-2.56}^{+3.16}$ & $686.40_{-254.80}^{+325.40}$ & $114.70_{-46.63}^{+57.61}$ \\
GRB~050802 & $1.7102$ & $70.40_{-11.50}^{+12.50}$ & $0.080$ & $9.04_{-1.54}^{+1.61}$ & $3.568$ & $14.90_{-4.08}^{+4.15}$ & $-0.410$ & $10.30_{-2.66}^{+2.71}$ & $62.19_{-10.19}^{+11.01}$ & $24.41_{-6.33}^{+6.45}$ \\
GRB~050803 & $0.422$ & $86.90_{-15.90}^{+21.30}$ & $82.000$ & $1.77_{-0.46}^{+0.65}$ & $80.456$ & $3.68_{-1.36}^{+1.65}$ & $82.688$ & $3.67_{-1.17}^{+1.50}$ & $5.45_{-1.00}^{+1.34}$ & $0.29_{-0.09}^{+0.12}$ \\
\enddata

\tablenotetext{a} {In units of $10^{-7}$~erg~cm$^{-2}$.}
\tablenotetext{b}{The start time of the time interval, when the peak count rate is reached,~s.}
\tablenotetext{c} {In units of $10^{-7}$~erg~cm$^{-2}$~s$^{-1}$}.
\tablenotetext{d} {In units of $10^{51}$~erg.}
\tablenotetext{e} {In units of $10^{51}$~erg~s$^{-1}$.}
\tablecomments{This table is available in its entirety in machine-readable form.}
\end{deluxetable}

\begin{deluxetable}{lccc r@{$\pm$}l r@{$\pm$}l rr}
\tabletypesize{\tiny} 
\rotate
\tablewidth{0pt}
\tablecaption{Collimation-corrected parameters\label{collimationtab}}
\tablehead {
\colhead{Burst}
& \colhead{$t_\textrm{jet}$}
& \colhead{CBM\tablenotemark{a}}
& \colhead{Ref.}
& \multicolumn{2}{c}{$\theta_\textrm{jet}$}
& \multicolumn{2}{c}{Collimation}
& \colhead{$E_{\gamma}$}
& \colhead{$L_{\gamma}$}
\\
\colhead{name}
& \colhead{(days)}
& \colhead{type}
& \colhead{}
& \multicolumn{2}{c}{(deg)}
& \multicolumn{2}{c}{factor ($\times 10^{-3}$)}
& \colhead{($10^{49}$~erg)}
& \colhead{($10^{49}$~erg~s$^{-1}$)}
}
\startdata
GRB~050730 & $0.17 \pm 0.02$ & HM & (\ref{En:Wang2018}) & $1.44$ & $0.07$ & $0.31$ & $0.03$ & $21.57_{-8.29}^{+10.45}$ & $3.60_{-1.51}^{+1.85}$ \\
GRB~050802 & $0.63$ & HM & (\ref{En:Kann2010}) & $3.98$ & $0.22$ & $2.41$ & $0.27$ & $14.98_{-2.98}^{+3.15}$ & $5.88_{-1.66}^{+1.69}$ \\
 & & SWM & & $3.40$ & $0.16$ & $1.76$ & $0.17$ & $10.93_{-2.07}^{+2.20}$ & $4.29_{-1.18}^{+1.20}$ \\
GRB~050904 & $2.6 \pm 1.0$ & HM & (\ref{En:Frail2006}) & $3.14$ & $0.32$ & $1.50$ & $0.32$ & $222.10_{-55.60}^{+56.90}$ & $25.70_{-7.89}^{+8.01}$ \\
GRB~051111 & $0.03 \pm 0.01$ & HM & (\ref{En:Wang2018}) & $1.26$ & $0.11$ & $0.24$ & $0.04$ & $1.91_{-0.40}^{+0.41}$ & $0.43_{-0.10}^{+0.10}$ \\
GRB~060206 & $1.83 \pm 0.09$ & HM & (\ref{En:Wang2018}) & $4.78$ & $0.07$ & $3.48$ & $0.11$ & $18.78_{-1.69}^{+2.08}$ & $21.73_{-2.35}^{+2.83}$ \\
 & & SWM & & $3.93$ & $0.07$ & $2.35$ & $0.09$ & $12.70_{-1.18}^{+1.43}$ & $14.69_{-1.62}^{+1.94}$ \\
GRB~060418 & $9.49 \pm 0.37$ & HM & (\ref{En:Wang2018}) & $10.20$ & $0.12$ & $15.80$ & $0.37$ & $232.40_{-14.51}^{+15.18}$ & $29.04_{-3.90}^{+3.94}$ \\
 & & SWM & & $5.51$ & $0.07$ & $4.62$ & $0.12$ & $67.99_{-4.30}^{+4.49}$ & $8.50_{-1.14}^{+1.16}$ \\
GRB~060526 & $1.41 \pm 0.14$ & HM & (\ref{En:Wang2018}) & $4.80$ & $0.18$ & $3.51$ & $0.27$ & $14.35_{-3.62}^{+5.12}$ & $24.25_{-6.46}^{+8.89}$ \\
 & & SWM & & $4.13$ & $0.23$ & $2.60$ & $0.30$ & $10.61_{-2.83}^{+3.89}$ & $17.93_{-5.01}^{+6.75}$ \\
GRB~060729 & $0.95 \pm 0.14$ & HM & (\ref{En:Wang2018}) & $7.33$ & $0.33$ & $8.17$ & $0.76$ & $7.13_{-2.31}^{+1.56}$ & $0.79_{-0.33}^{+0.27}$ \\
 & & SWM & & $7.08$ & $0.37$ & $7.62$ & $0.83$ & $6.65_{-2.19}^{+1.51}$ & $0.74_{-0.31}^{+0.26}$ \\
GRB~070411 & $1.76 \pm 0.86$ & HM & (\ref{En:Wang2018}) & $4.53$ & $0.60$ & $3.12$ & $0.87$ & $48.55_{-17.18}^{+20.08}$ & $8.13_{-3.23}^{+3.48}$ \\
 & & SWM & & $3.18$ & $0.31$ & $1.54$ & $0.32$ & $23.92_{-7.13}^{+8.78}$ & $4.01_{-1.39}^{+1.53}$ \\
GRB~080310 & $0.34 \pm 0.04$ & HM & (\ref{En:Wang2018}) & $3.08$ & $0.10$ & $1.45$ & $0.10$ & $5.36_{-0.79}^{+0.90}$ & $1.50_{-0.31}^{+0.31}$ \\
 & & SWM & & $3.13$ & $0.10$ & $1.49$ & $0.10$ & $5.50_{-0.81}^{+0.92}$ & $1.54_{-0.32}^{+0.32}$ \\
GRB~081008 & $0.11 \pm 0.01$ & HM & (\ref{En:Wang2018}) & $1.80$ & $0.05$ & $0.49$ & $0.03$ & $7.11_{-1.30}^{+1.52}$ & $0.76_{-0.17}^{+0.19}$ \\
GRB~090426 & $0.29 \pm 0.07$ & HM & (\ref{En:Wang2018}) & $4.01$ & $0.27$ & $2.45$ & $0.34$ & $0.59_{-0.14}^{+0.15}$ & $3.86_{-0.98}^{+1.04}$ \\
GRB~091029 & $0.35 \pm 0.08$ & HM & (\ref{En:Wang2018}) & $2.52$ & $0.16$ & $0.96$ & $0.12$ & $15.10_{-3.08}^{+3.02}$ & $3.30_{-0.76}^{+0.75}$ \\
GRB~110205A & $1.19 \pm 0.09$ & HM & (\ref{En:Wang2018}) & $3.54$ & $0.07$ & $1.91$ & $0.08$ & $120.40_{-10.00}^{+10.62}$ & $7.62_{-0.93}^{+0.98}$ \\
 & & SWM & & $2.14$ & $0.04$ & $0.70$ & $0.03$ & $43.83_{-3.54}^{+3.77}$ & $2.77_{-0.33}^{+0.35}$ \\
\enddata
\tablerefs{\cititem{Wang2018}; \cititem{Kann2010}; \cititem{Frail2006}; }

\tablenotetext{a}{In cases where the preferred CBM density profile is known, it is taken from the same paper as $t_\textrm{jet}$. Otherwise, calculations for both CBMs are provided and the geometric mean of the HM and WM energetics is used in the calculations.}
\end{deluxetable}

\begin{deluxetable}{lcccccccc}
\tabletypesize{\scriptsize} 
\tablewidth{0pt}
\tablecaption{Hardness-intensity correlations\label{correlationtab}}
\tablecolumns{9}
\tablehead{
\colhead{Correlation}
& \colhead{$N$}
& \colhead{$\rho_S$}
& \colhead{$P_{\rho_S}$}
& \colhead{$a$}
& \colhead{$b$}
& \colhead{$a_{\sigma_\textrm{int}}$}
& \colhead{$b_{\sigma_\textrm{int}}$}
& \colhead{$\sigma_\textrm{int}$}
\\
\colhead{}
& \colhead{}
& \colhead{}
& \colhead{}
& \colhead{}
& \colhead{}
& \colhead{} 
& \colhead{}
& \colhead{}
}
\startdata
$E_\textrm{p,i}$ versus $S$ & $315$ & $0.64$ & $< 0.00001$ & $0.382 \pm 0.002$ & $3.89 \pm 0.01$ & $0.301 \pm 0.017$ & $3.66 \pm 0.09$ & $0.217$\\
$E_\textrm{p,p}$ versus $F_\textrm{peak}$ & $315$ & $0.59$ & $< 0.00001$ & $0.398 \pm 0.003$ & $4.41 \pm 0.02$ & $0.304 \pm 0.019$ & $4.01 \pm 0.11$ & $0.234$\\
$E_\textrm{p,i,z}$ versus $E_\textrm{iso}$ & $315$ & $0.7$ & $< 0.00001$ & $0.429 \pm 0.002$ & $-20.18 \pm 0.12$ & $0.307 \pm 0.016$ & $-13.61 \pm 0.87$ & $0.225$\\
$E_\textrm{p,p,z}$ versus $L_\textrm{iso}$ & $315$ & $0.7$ & $< 0.00001$ & $0.429 \pm 0.004$ & $-19.87 \pm 0.19$ & $0.310 \pm 0.017$ & $-13.59 \pm 0.90$ & $0.235$\\
$E_\textrm{p,i,z}$ versus $E_{\gamma}$ & $43$ & $0.57$ & $0.00006$ & $0.566 \pm 0.008$ & $-26.01 \pm 0.39$ & $0.388 \pm 0.066$ & $-16.98 \pm 3.36$ & $0.272$\\
$E_\textrm{p,p,z}$ versus $L_{\gamma}$ & $43$ & $0.43$ & $0.004$ & $0.679 \pm 0.014$ & $-31.24 \pm 0.70$ & $0.269 \pm 0.082$ & $-10.71 \pm 4.11$ & $0.351$\\
\enddata

\tablecomments{$N$ is the number of bursts in the fit sample, $\rho_S$ is a Spearman rank-order correlation coefficient, $P_{\rho_S}$ is the corresponding chance probability, $a$ ($a_{\sigma_\textrm{int}}$) and $b$ ($b_{\sigma_\textrm{int}}$) are the slope and the intercept for the fits without (with) intrinsic scatter $\sigma_\textrm{int}$. }
\end{deluxetable}

\begin{deluxetable}{lccccc}
\tabletypesize{\scriptsize}
\tablewidth{0pt}
\tablecaption{LF and EF fits with BPL and Cutoff PL\label{tabLFEF}}
\tablehead {
\colhead{Data}
& \colhead{Evolution}
& \colhead{Model}
& \colhead{$\alpha_1$}
& \colhead{$\alpha_2$}
& \colhead{$\textrm{log}$~$x_{b,52}$}
\\
\colhead{}
& \colhead{(PL index)}
& \colhead{}
& \colhead{}
& \colhead{}
& \colhead{($\textrm{log}$~$x_\textrm{cut,52}$)}
}

\startdata
$\psi(L')            $ & $\delta_L$=1.2 & BPL & $-0.28_{-0.07}^{+0.04}$ & $-1.01_{-0.28}^{+0.10}$ & $0.21_{-0.18}^{+0.28}$\\
$\psi(L')            $ & $\delta_L$=1.2 & CPL & $-0.45_{-0.04}^{+0.08}$ &      \nodata            & $1.67_{-0.49}^{+0.19}$\\
$\psi(E')            $ & $\delta_E$=1.1 & BPL & $-0.29_{-0.09}^{+0.01}$ & $-1.01_{-0.44}^{+0.02}$ & $0.80_{-0.06}^{+0.44}$\\
$\psi(E')            $ & $\delta_E$=1.1 & CPL & $-0.35_{-0.04}^{+0.03}$ &      \nodata            & $1.74_{-0.11}^{+0.07}$\\
\hline
$\psi(L_\textrm{iso})$ & no evolution   & BPL & $-0.26_{-0.05}^{+0.02}$ & $-0.99_{-0.17}^{+0.05}$ & $0.78_{-0.08}^{+0.17}$\\
$\psi(L_\textrm{iso})$ & no evolution   & CPL & $-0.38_{-0.03}^{+0.04}$ &      \nodata            & $2.02_{-0.19}^{+0.11}$\\
$\psi(E_\textrm{iso})$ & no evolution   & BPL & $-0.26_{-0.08}^{+0.01}$ & $-0.97_{-0.44}^{+0.03}$ & $1.24_{-0.08}^{+0.46}$\\
$\psi(E_\textrm{iso})$ & no evolution   & CPL & $-0.32_{-0.04}^{+0.02}$ &      \nodata            & $2.21_{-0.09}^{+0.06}$\\
\enddata

\end{deluxetable}
\end{document}